\documentclass[reprint, nofootinbib, amsmath,amssymb, aps]{revtex4-1}

\usepackage[colorlinks=true,citecolor=blue,linkcolor=blue,urlcolor=blue, backref=false,pdfborder={0 0 0}]{hyperref}

\usepackage{aas_macros} 
\usepackage{soul}
\usepackage{tablefootnote}

\newcommand{\threej}[6]{{\begin{pmatrix}
#1 & #2 & #3 \\
#4 & #5 & #6
\end{pmatrix}}}

\newcommand{\Qgeo}[5]{{Q}_{ #1\ell_{#2} \ell_{#3},\ell_{#4}\ell_{#5}}^{m_{#2} m_{#3}, m_{#4} m_{#5}}}

\newcommand{\deriv}[2]{\frac{\textrm{d} #1}{\textrm{d} #2}}

\newcommand{\beq}{\begin{equation}}
\newcommand{\eeq}{\end{equation}}
\newcommand{\barr}{\begin{eqnarray}}
\newcommand{\earr}{\end{eqnarray}}
\newcommand{\ball}{\begin{align}}
\newcommand{\eall}{\end{align}}
\newcommand{\bs}{\boldsymbol}

\newcommand{\hn}{\hat{n}}
\newcommand{\hk}{\hat{k}}
\newcommand{\bx}{\bs{x}}
\newcommand{\fbh}{f_{\rm pbh}}
\newcommand{\lsim}{\mathrel{\hbox{\rlap{\lower.55ex\hbox{$\sim$}} \kern-.3em \raise.4ex \hbox{$<$}}}}
\newcommand{\gsim}{\mathrel{\hbox{\rlap{\lower.55ex\hbox{$\sim$}} \kern-.3em \raise.4ex \hbox{$>$}}}}
\usepackage{cancel}
\newcommand{\sld}{\cancel{\delta}}
\newcommand{\bk}{\boldsymbol{k}}

\newcommand{\av}[1]{\langle{#1}\rangle}

\usepackage{enumitem}
\usepackage{cancel}
\usepackage{graphicx}
\usepackage{dcolumn}
\usepackage{bm}

\usepackage{braket}
\usepackage{placeins}
\usepackage{color,soul}
\usepackage[toc,page]{appendix}

\begin{document}
\title{CMB polarization non-Gaussianity from accreting primordial black holes}
\author{Trey W. Jensen}
\author{Yacine Ali-Ha{\"i}moud}
\affiliation{Center for Cosmology and Particle Physics \\
Department of Physics, New York University \\
New York, NY 10003, USA}

\date{\today}

\begin{abstract}
Primordial black holes (PBHs) would induce non-Gaussianity in the cosmic microwave background (CMB) by sourcing recombination perturbations spatially modulated by relative velocities between PBHs and the baryons they accrete. The leading non-Gaussian signatures are non-vanishing connected 4-point correlation functions, or trispectra. Earlier, we computed the CMB temperature trispectrum, and forecasted Planck to be more sensitive to it than to changes in the CMB temperature power spectrum for light enough PBHs. Excitingly, accreting PBHs would also induce non-Gaussianity in CMB polarization, and source both $E$ and $B$ modes, which we compute in this paper. We first calculate linear-response perturbations to the tensor-valued photon distribution function sourced by a general spatially-varying ionization history, and apply our results to accreting PBHs. We then compute linear-order perturbations to the temperature and polarization 2-point functions sourced by inhomogeneities in recombination due to accreting PBHs; we find them to be negligible relative to their counterparts sourced by homogeneous perturbations to the ionization history. Lastly, we compute all CMB trispectra including temperature, $E$- and $B$-mode polarization at linear order in the PBH abundance. We forecast that including polarization data in a 4-point-function analysis would only increase Planck's sensitivity to accreting PBHs by a factor $\sim 2$ relative to using temperature alone. As a consequence, we find that a search for PBHs using all temperature and polarization trispectra with Planck data would mostly not be competitive with current bounds from temperature and polarization power spectra. In contrast, we forecast that a CMB Stage-4 experiment would gain significant sensitivity to accreting PBHs through a 4-point-function search, in particular through the contributions of parity-odd trispectra including one $B$-mode field.
\end{abstract}

\maketitle

\section{Introduction}\label{sec:intro}
Primordial black holes (PBHs) could have formed during the radiation-dominated era due to the collapse of large-amplitude small-scale density fluctuations \citep{hawking71a}. The time and scales at which these over-densities collapse directly relate to the mass of the PBHs, and could theoretically span many orders of magnitudes (for reviews see Refs.~\citep{carr20a, carr21a}). PBHs are a historical dark matter candidate, and have moreover been invoked as a possible explanation for a variety of enigmas: they could be the seed for supermassive black holes \cite{bean02a,bernal18a}, produce the Galactic $\gamma$--ray background \citep{carr16b}, or account for recent LIGO/Virgo gravitational wave observations \citep{bird16a}. Even if they only contribute a small fraction to the dark matter mass budget, a proof of their existence -- or strong upper limits to their abundance -- would at the very least shine light on the physics present in the \textit{very} early Universe, on scales much smaller than those directly accessible to linear-cosmology observables.

In this paper we concern ourselves with the intermediate-mass range of PBHs ($\sim 1$--$10^4$ $M_{\odot}$), which is most strongly constrained by a null signal in the cosmic microwave background (CMB). In this mass range, PBHs would accrete matter from the surrounding gas, and the radiation powered by this accretion would then deposit energy back into the cosmological plasma. This additional energy source would delay the primordial plasma's recombination, consequently altering the CMB last-scattering surface and ultimately the statistics of CMB photons observed today. It is through this phenomenon, and specifically through its effect on CMB anisotropy angular power spectra (or 2-point correlation functions), that some of tightest constraints were derived on stellar-to-intermediate-mass PBHs \citep{ricotti08a, yacine17a,poulin17a}. 

However, the story does not end at 2-point statistics. Indeed, as uncertain as the PBH accretion efficiency and radiation power may be, they almost certainly depend on the local relative velocity between PBHs and the baryons they accrete, in addition to the background baryon sound speed \citep{ricotti08a, yacine17a,poulin17a}. Under the assumption that PBHs trace adiabatic cold dark matter (CDM) perturbations on large scales, and given that small-scale non-linear velocities resulting from PBH clustering are subdominant to their large-scale counterparts \cite{inman19}, this implies that the radiative energy output of PBHs is a function of the (large-scale) relative velocity $v_{\rm bc}$ between baryons and CDM. It is well-known that these relative velocities are significantly supersonic around recombination and fluctuate on $\sim 100$ Mpc scales \cite{Tseliakhovich_10}. As a consequence, we expect the PBH accretion-powered luminosity to be \emph{significantly} modulated on large scales, resulting in inhomogeneous perturbations to the cosmic free-electron fraction, which would then generate non-Gaussian signatures in CMB anisotropies \citep{senatore09a,khatri09a, dvorkin13a, jensen23a}. Interestingly, in contrast with the more commonly studied CMB bispectrum (or 3-point function), sourced e.g.~by inhomogeneously annihilating DM \cite{dvorkin13a}, the lowest-order non-Gaussianity induced by PBHs is the trispectrum (or 4-point function), due to the dependence of PBH accretion luminosity on $v_{\rm bc}^2$. The present work is the last of a series of three papers quantifying these \emph{qualitatively novel} non-Gaussian signatures of accreting PBHs and their detectability.

In Ref.~\citep{jensen21a} (hereafter Paper~I), we explicitly computed the spatial perturbations to the free-electron fraction, carefully accounting for the finite propagation of injected photons before they deposit their energy into the plasma with radiation transport simulations. Our main result was that the inhomogeneous perturbation to the free-electron fraction $\Delta x_e^{\rm inh}$ is comparable in magnitude to its homogeneous counterpart $\Delta \overline{x}_e$. Denoting by $X$ the CMB temperature or polarization anisotropy, this result implies that the non-Gaussianity in CMB anisotropies $X^{\rm NG}$ sourced by $\Delta x_e^{\rm inh}$ should be comparable to the perturbation to the Gaussian signal $\Delta X^{\rm G}$ sourced by $\Delta \overline{x}_e$. This preliminary result was promising. Indeed, the Planck satellite is sensitive to sub-percent level changes to the Gaussian part of the CMB signal, $|\Delta X^{\rm G}| \lesssim (10^{-3}-10^{-2}) |X^{\rm G}|$ \cite{planckbluebook}. On the other hand, current CMB limits on primordial non-Gaussianity \cite{planck20c} can be approximately translated to $|X^{\rm NG}| \lesssim (10^{-4}-10^{-3}) |X^{\rm G}|$, depending on their specific type and shape. We may therefore expect, a priori, that the non-Gaussian signal could imply an improvement in sensitivity to accreting PBHs over the CMB 2-point functions limits by a factor of $\sim 1-10^2$. In light of this expectation, in Ref.~\cite{jensen23a} (hereafter Paper~II) we set out to compute the CMB temperature trispectrum sourced by inhomogeneously-accreting PBHs, and forecasted Planck's sensitivity to this signal. We found that the constraining power of the temperature 4-point function $\av{TTTT}$ is comparable to that of the temperature 2-point function $\av{TT}$, with the former becoming more sensitive than the latter for PBH masses below $\sim 10^3 M_{\odot}$. 

Given that CMB-anisotropy power spectra limits on accreting PBHs are actually dominated by $E$-mode polarization cross- and auto-power spectra $\av{TE}$ and $\av{EE}$, one may expect that the same holds true for trispectra. Moreover, while the forecasted improvement in sensitivity from temperature trispectra is at the lower end of our order-of-magnitude estimates, it is still possible, in principle, that the 4-point functions involving $E$-mode polarization, namely $\av{TTTE}, \av{TTEE}, \av{TEEE}$ and $\av{EEEE}$, may be more significantly constraining than their 2-point function counterparts. Last but not least, inhomogeneous recombination would also source non-Gaussian $B$ modes, leading to non-vanishing parity-odd trispectra $\av{TTTB}, \av{TTEB}, \av{TEEB}$ and $\av{EEEB}$ at linear order in the PBH abundance. These trispectra are especially interesting as, absent primordial tensor perturbations, the cosmic variance in $B$ modes is much reduced relative to that in $E$ modes, as it only arises from lensing \cite{Meerburg_16}. With these auspicious motivations in mind, the goal of this paper is to compute all temperature-polarization trispectra sourced by accreting PBHs, and extend the forecasts done in Paper~II to include polarization data.

The remainder of this paper is organized as follows. Section~\ref{sec:paperI} briefly reviews how Paper~I and Paper~II compute a free-electron fraction perturbation from an accreting PBH energy source and the approximations therein to reduce the computational cost of calculating trispectra. In Section~\ref{sec:temp_ani}, we derive the standard expressions for $E$- and $B$-mode polarization starting from geometric arguments on a tensor-valued photon distribution function, and generalize them when introducing an inhomogeneous perturbation to the free-electron fraction. We apply this formalism to accreting PBHs: in Section~\ref{sec:powerspec} we compute the perturbed power spectra, and in Section~\ref{sec:trispec} we arrive at the main result of this work, namely all the induced trispectra incorporating all three $T$, $E$, and $B$ modes. In Section~\ref{sec:forecast}, we forecast the sensitivity of Planck and of a CMB Stage-4-like experiment to these novel trispectra and discuss the results. Finally, we conclude in Section~\ref{sec:conc}.

\section{Perturbed recombination from accreting PBHs}\label{sec:paperI}

In this section we briefly review the effect of accreting PBHs on the ionization history. As in Papers I and II, we use the prescription of PBH luminosity from Ref.~\cite{yacine17a} (hereafter AK17), specifically the ``collisional ionization" case, which is the most conservative (see AK17 for details). We use the results of Paper I for the inhomogeneous ionization-fraction perturbation due to PBHs, and use the same approximate factorized form of the free-electron fraction as in Paper~II. This factorization greatly reduces the computational cost of calculating the trispectra induced by accreting PBHs. From here on we denote $D(k_1 \cdots k_N) \equiv d^3 k_1/(2 \pi)^3  \cdots ~ d^3 k_N/(2 \pi)^3$ and $\sld(\bk) \equiv (2 \pi)^3 \delta_{\rm D}(\bk)$. 

\subsection{General equations}

If PBHs in the mass range $\sim 1$--$10^4$ $M_{\odot}$ were to exist in the early Universe, they would accrete, compress, heat, and ionize baryons eventually causing the accreted gas to emit free-free radiation. As emphasized in AK17, the PBH accretion efficiency, and thus their luminosity $L$, is dependent on the local relative velocity between the baryons and the cold dark matter ${\bm v}_{\rm bc}({\bm r})$. When this relative velocity is larger than the baryonic speed of sound, which is typically the case around recombination \cite{Tseliakhovich_10}, the accretion rate is heavily suppressed. As a consequence, the PBH luminosity $L(v_{\rm bc}(\bs{r}))$ is strongly inhomogeneous due to the large-scale fluctuations of ${\bm v}_{\rm bc}({\bm r})$.

We write a general free-electron fraction as $x_e(\bs{r})= x_e^{(0)}+\Delta x_e(\bs{r})$, where $x_e^{(0)}$ is the standard, homogeneous $\Lambda$CDM free-electron fraction, and $\Delta x_e(\bs{r})$ is the perturbation sourced by accreting PBHs. Note that $\Delta x_e$ has both a homogeneous part $\Delta \overline{x}_e$ and an inhomogeneous part $\Delta x_e^{\rm inh}(\bs{r}) = \Delta x_e(\bs{r}) - \Delta \overline{x}_e$. Under the assumption that free-electron perturbations are small and respond approximately linearly to energy injection, one may obtain $\Delta x_e$ from the volumetric energy injection rate $\dot{\rho}_{\rm inj}$ with a Green's function. Explicitly, in Fourier space,
\begin{align}
\Delta x_e^{\rm inh}(z, \bs{k}) &= \int_z^{\infty} \frac{d z'}{1+z'}  G_{x_e}^{\rm inj}(z, z', k) \nonumber\\
&~~~~~~~ \times \frac{\overline{\dot{\rho}}_{\rm inj}}{n_{\rm H} H E_I}\Big{|}_{z'} \delta_{\rm inj}(z', \bk),\label{eq:Dxe(k)}
\end{align}
where $z'$ is the injection redshift, $n_{\rm H}$ is the mean number density of hydrogen, $H$ is the Hubble rate, $E_I \equiv 13.6$ eV is hydrogen's ionization energy, $\overline{\dot{\rho}}_{\rm inj}$ is the spatially averaged volumetric photon energy injection rate and $\delta_{\rm inj}$ is its fractional spatial fluctuation. For accreting PBHs with mass $M_{\rm pbh}$ making a fraction $f_{\rm pbh}$ of dark matter, the mean volumetric energy injection rate is given by
\begin{align}
\overline{\dot{\rho}}_{\rm inj}(z) &\equiv f_{\rm pbh}\frac{\overline{\rho}_c(z)}{M_{\rm pbh}} \overline{L}(z),
\end{align}
where $\overline{\rho}_c$ is the mean dark matter mass density and $\overline{L}(z)$ is the spatially-averaged (or equivalently, relative-velocity-averaged) PBH luminosity as computed in AK17. Note that this equation assumes a Dirac-delta mass function for PBHs, and is trivially generalizable to an extended mass distribution. The relative perturbation in energy injection is simply given by the relative fluctuation of PBH luminosity, $\delta_{\rm inj}(\bs{r}) = L(\bs{r})/\overline{L} -1$.

The homogeneous perturbation $\Delta \overline{x}_e(z)$ is obtained with an equation similar to Eq.~\eqref{eq:Dxe(k)}, with the substitutions $G_{x_e}^{\rm inj}(z, z', k) \rightarrow G_{x_e}^{\rm inj}(z, z', k = 0)$ and $\delta_{\rm inj}(z,k)\rightarrow1$. 

 The result of Paper I was to evaluate $G_{x_e}^{\rm inj}(z, z', k)$ for injection of $\sim 0.1-10$ MeV photons, by combining a radiation transport code with a modified \texttt{HYREC}-2 \cite{hyrec2, yacine11a, YAH_10}, thereby generalizing calculations of the spatially-averaged Green's function $G_{x_e}^{\rm inj}(z, z', k = 0)$ which can be found, e.g.~in Refs.~\cite{slatyer16a,hongwan20a}.

\subsection{Approximate factorized form}

In principle we should obtain $\delta_{\rm inj}(\bs{k})$ from Fourier-transforming $\delta_{\rm inj}(\bs{r}) = L(v_{\rm bc}(\bs{r}))/\overline{L} -1$. For simplicity, and given the large theoretical uncertainty underlying $L(v_{\rm bc})$ anyway, we approximate $\delta_{\rm inj}$ as a biased tracer of $v_{\rm bc}^2$:
\beq\label{eq:b}
\delta_{\rm inj}(z,\bs{r}) \approx b(z) \left(\frac{v_{\rm bc}^2(z,\bs{r})}{\langle v_{\rm bc}^2 \rangle(z)} -1 \right), \ \ \ b \equiv \frac32 \frac{\langle v_{\rm bc}^2 \delta_{\rm inj}\rangle}{\langle v_{\rm bc}^2\rangle}.
\eeq
We will confirm later in Section~\ref{sec:temp_ani} that this approximation is in fact exact at the level of 2-point correlation functions, and a good first step for cross-correlations relevant to 4-point function calculations. With this approximation, the energy injection rate, hence the ionization perturbation, are quadratic in the primordial curvature perturbation $\zeta$. Denoting the relative free-electron fraction perturbation by $\delta_e \equiv \Delta x_e/x_e^{(0)} = \overline{\delta}_e + \delta_e^{\rm inh}$, we have, for $\bk \neq 0$, 
\begin{align}
\delta_e^{\rm inh}(z, \bk) &\approx \fbh \int\!\! D(k_1 k_2)\sld(\bk_1 + \bk_2 - \bk) \nonumber\\
&\quad\times T_e(z, \bk_1, \bk_2)  \zeta(\bk_1) \zeta(\bk_2),  \label{eq:Te-def}
\end{align}
where $T_e(z, \bk_1, \bk_2)$ is given explicitly in Paper II, and is a redshift integral whose integrand is proportional to $G_{x_e}^{\rm inj}(z, z',|\bk_1 + \bk_2|)$ and quadratic in the relative velocity transfer function $\widetilde{v}_{bc}(z', k)$, defined through $\bs{v}_{\rm bc}(z, \bk) = - i \hk \widetilde{v}_{bc}(z', k) \zeta(\bk)$. 

With the motivation of reducing the computational load in obtaining trispectra, in Paper~II we derived an approximate factorized form for $T_e(z, \bk_1, \bk_2)$, by making the following assumptions. First, we approximated the scale-dependence of the relative velocity, $v_{\rm bc}(k)/\langle v_{\rm bc}^2\rangle ^{1/2}$, as constant in redshift. This is an accurate approximation after decoupling at $z \lesssim z_{\rm dec}\approx 1020$, at which point $v_{\rm bc}\propto (1+z)$ regardless of scale \citep{Tseliakhovich_10}. Moreover, this approximation turns out to be accurate even before decoupling because $G_{x_e}^{\rm inj}(z,z',k)$ is peaked at $z' \approx z$ for $z\gsim 10^3$ (c.f. Fig.~9 in Paper~I). Second, we approximated the remaining redshift integral of the energy-injection-weighted Green's function by a factorized form that bounds it from below. In Paper II we explicitly showed that considering instead the limit of spatially on-the-spot energy deposition, which bounds the Green's function from above, results in differences below $\sim 20\%$, far below the theoretical uncertainty arising from the accretion problem. 

Ultimately, with these approximations, and as in Paper~II, we approximate the ionization transfer function defined in Eq.~\eqref{eq:Te-def} as
\begin{align}
T_e(z, \bk_1,\bk_2) &\approx -(\hat{k}_1 \cdot \hat{k}_2) \Delta_e(z, k_1) \Delta_e(z, k_2), \label{eq:dele_pbh}\\
\Delta_e(z, k) &\equiv \frac{G_e(z, \sqrt{2} k)}{\sqrt{G_e(z, 0)}} \frac{\widetilde{v}_{\rm bc}(z, k)}{\langle v_{\rm bc}^2 \rangle_z^{1/2}}, \label{eq:Delta_e-def}
\end{align}
where $G_e$ is the energy-injection-weighted Green's function defined as
\begin{align}
G_e(z, k) \equiv& \int_z^{\infty}\frac{d z'}{1+z'} \frac{G_{x_e}^{\rm inj}(z, z', k)}{x_e^{(0)}(z)} \frac{\overline{\rho}_c \overline{L}~ b}{M_{\rm pbh} n_{\rm H} H E_I}\Big{|}_{z'}.\label{eq:G_e}
\end{align}
This approximation may be reformulated as follows. If we define the longitudinal vector field $\bs{u}$ with Fourier components $\bs{u}(z, \bk) = - i \hk ~ (\Delta_e(z, k)/\sqrt{G_e(z, 0)}) \zeta(\bk)$, then Eqs.~\eqref{eq:Te-def} and \eqref{eq:dele_pbh} may be simply rewritten as the following local expression for the free-electron perturbation:
\beq
\delta_e^{\rm inh}(z, \bs{r}) \approx f_{\rm pbh} \overline{G}_e(z) ~\left( u^2(z, \bs{r}) - \av{u^2}(z) \right), \label{eq:delta_e-u^2}
\eeq
where $\overline{G}_e(z) \equiv G_e(z, k = 0)$. 

Before ending this section, let us emphasize that Eq.~\eqref{eq:Te-def} only holds for $\bk \neq 0$. Indeed, one should subtract a term proportional to $\sld(\bk)$ to account for the second term in Eq.~\eqref{eq:b}, and to enforce that $\langle \delta_{\rm inj} \rangle = 0 = \langle \delta_e^{\rm inh} \rangle$. Equivalently, we may enforce that the spatial average of $\delta_e^{\rm inh}$ vanishes for any given realization (rather than enforcing that its cosmological average vanishes). This would amount to enforcing $\delta_e^{\rm inh} (\bk = 0) = 0$. To do so, we enforce that $T_e(z, \bk_1, - \bk_1) = 0$.

\section{Tensor-valued photon distribution function: general equations}\label{sec:temp_ani}

In this section we compute the solution to the tensor-valued distribution function from the Boltzmann-Einstein equations in the presence of an arbitrary deviation from the standard free-electron fraction evolution. We already addressed the evolution of total intensity perturbations in Paper II, and therefore focus on the photon polarization here. Given that Planck data is closely consistent with the standard (and homogeneous) recombination history \citep{planck20c}, we restrict ourselves to small (but generally spatially-dependent) perturbations to the free-electron fraction, allowing the use of perturbation theory. 

\subsection{Collisional Boltzmann equation for the polarization tensor}
Photons and their polarization can be succinctly described as a Hermitian tensor-valued distribution function $I_{ij}(\hat{n})$, which is transverse to the direction of propagation $\hn$. We we can then write this tensor as
\begin{align}\label{eq:intensity_tens}
    I_{ij}(\hn)=\frac{1}{2}(\delta_{ij}-n_i n_j)I(\hn)+\frac{1}{2}P_{ij}(\hn),
\end{align}
where $i,j$ run from 1 to 3, and we have split the tensor into a trace part $I$, and a symmetric traceless part $P_{ij}$ (we neglect circular polarization) also orthogonal to the direction of propagation $P_{ij}n^i=0$. The total intensity $I$ is related to the temperature perturbation $I\propto \overline{T}^3(1+\Theta)$, where $\overline{T}$ is the background CMB temperature. 

The evolution of this tensor-valued photon intensity is governed by the Boltzmann-Einstein differential system. CMB photons follow geodesics in an expanding universe subject to Thomson scattering off free electrons. Specifically, the Boltzmann equation for the polarization is 
\begin{align}\label{eq:be_ODE}
    \deriv{P_{ij}}{\eta}\equiv \dot{P}_{ij}+\hat{n}\cdot\bs{\nabla} P_{ij}&=\dot{\tau}\deriv{P_{ij}}{\tau}\Bigr|_{\rm Th},
\end{align}
where overdots denote partial derivatives with respect to conformal time, $\bs{\nabla}$ is the gradient with respect to comoving spatial coordinates, and $\dot{\tau}\equiv an_e\sigma_T$ is the conformal Thomson scattering rate. The Thomson scattering collision operator takes the following form
\begin{align}
\deriv{P_{ij}(\hn)}{\tau}\Bigr|_{\rm Th}=&\left(\delta_{ik}^{\perp \hn}\delta_{jl}^{\perp \hn}- \frac12\delta_{ij}^{\perp \hn}\delta_{kl}^{\perp \hn}\right)\Pi_{kl}-P_{ij}(\hn), \label{eq:dotP-Thoms}\\
\delta_{ij}^{\perp \hn}\equiv&~ \delta_{ij}-n_in_j,\quad\quad
\Pi_{kl}\equiv \frac32\langle P_{kl}\rangle + \sigma_{kl},
\end{align}
where the angled brackets denote an angular average, and
\begin{align}
    \sigma_{ij}\equiv \frac12 \int \frac{d^2 \hn}{4\pi}\left(\delta_{ij}-3n_in_j\right)\Theta(\hn)
\end{align}
is the real-space photon temperature quadrupole moment. Note that the symmetric trace-free tensor $\Pi_{ij}$ is independent of direction $\hn$, so that all the angular dependence in the first term of Eq.~\eqref{eq:dotP-Thoms} comes from the transverse-trace-free projection operator acting on $\Pi_{kl}$.

\subsection{Helicity basis and \texorpdfstring{$E/B$}{E/B} decomposition}

The polarization tensor can be described by its components in any given basis tangent to the 2-sphere. Introducing spherical polar coordinates and the associated local orthonormal basis ($\hn, \hat{e}_\theta, \hat{e}_\phi)$, we define the helicity basis 
\begin{align}\label{eq:helicity}
    \bs{\epsilon}_{\pm}(\hn)\equiv\frac{1}{\sqrt{2}}(\hat{e}_\theta \pm i \hat{e}_\phi),
\end{align}
from which we obtain the spin-2 quantities\footnote{The mapping with the Stokes parameters $Q, U$ is $P_{\pm} = -(Q \pm i U)$.} \citep{lewis06a} 
\begin{align}
    P_{\pm}(\hn) \equiv -\epsilon_{\pm}^i \epsilon_{\pm}^j P_{ij}(\hn).
\end{align}
The Boltzmann equation \eqref{eq:be_ODE} does not contain any angular derivative operators and can therefore be readily projected on the helicity basis. Using the fact that $\epsilon_{\pm}^i n^i = 0 = \epsilon_{\pm}^i \epsilon_{\pm}^i$ when projecting the Thomson scattering collision operator \eqref{eq:dotP-Thoms}, we arrive at the simple equation for the two helicities:
\beq
\dot{P}_{\pm} + \hn \cdot \bs{\nabla} P_{\pm} = \dot{\tau} \left(-\epsilon_{\pm}^i \epsilon_{\pm}^j \Pi_{ij} - P_{\pm}\right). \label{eq:CBE P_pm}
\eeq

The helicity basis is useful because spin-2 fields can be decomposed with spin-weighted spherical harmonics to compute the familiar $E$ and $B$ polarizations,
\begin{align}\label{eq:E&B_decomp}
    E_{\ell m}\pm iB_{\ell m}=\int d^2\hn~ Y^{\pm 2*}_{\ell m}(\hn)P_{\pm}(\hn) \equiv P^{\pm}_{\ell m},
\end{align}
where $Y^{\pm 2}_{\ell m}$ are spin-weighted spherical harmonics reviewed briefly in Appendix~\ref{app:spin}.

\subsection{Standard solution}\label{sec:canon}
In this section we review the canonical solution to the Boltzmann equation for polarization with scalar initial conditions, linear evolution, and standard, homogeneous recombination. We denote all variables in this standard setup by a superscript $(0)$, e.g.~$\dot{\tau}^{(0)}$ is the standard, homogeneous differential Thomson optical depth, and $P_{\pm}^{(0)}$ are the standard helicity components of the polarization tensor. In the standard scenario, the Fourier components of any scalar field $X^{(0)}(\bk, \hn)$ may depend on direction only through $\mu \equiv \hn \cdot \hk$. As a consequence, we may expand any such field on basis of Legendre polynomials $P_\ell(\mu)$:
\beq
X^{(0)}(\eta, \bk, \hn)=\sum_\ell (-i)^\ell (2\ell+1) X_\ell^{(0)}(\eta, \bk) P_\ell(\mu). \label{eq:Legendre expansion}
\eeq
For scalar initial conditions, the symmetric-trace-free tensor $\Pi_{ij}$ takes the form 
\begin{align}
    \Pi^{(0)}_{ij}(\eta, \bk)&= \frac12 (3\hk_i\hk_j-\delta_{ij})\Pi^{(0)}(\eta, \bk), \\
    \Pi^{(0)} &\equiv \Delta^{(0)}_{P0}+\Delta^{(0)}_{P2}+\Theta^{(0)}_2,
\end{align}
where $\Delta_P\equiv \frac{\hk_i \hk_j}{1-(\hk \cdot \hn)^2}P_{ij}$, $\Delta_{P \ell}^{(0)}$ are its Legendre-expansion coefficients, and $\Theta_2^{(0)}$ is the $\ell = 2$ coefficient (i.e.~quadrupole) of the photon temperature field. Since $\epsilon_{\pm}^i \epsilon_{\pm}^i = 0$, we then have 
\beq
\epsilon_{\pm}^i \epsilon_{\pm}^j \Pi_{ij}^{(0)}(\eta, \bk) = \frac32 (\bs{\epsilon}_{\pm}\cdot \hk)^2 \Pi^{(0)}(\eta, \bk). \label{eq:Pi_ij^0}
\eeq
In Fourier space, the Boltzmann equation for $P_{\pm}^{(0)}$, Eq.~\eqref{eq:CBE P_pm}, then becomes the simple ordinary differential equation (ODE)
\beq
\dot{P}_{\pm}^{(0)} + i (\hn \cdot \bk) P_{\pm}^{(0)} + \dot{\tau}^{(0)} P_{\pm}^{(0)} = -\frac32 \dot{\tau}^{(0)} (\bs{\epsilon}_{\pm}\cdot \hk)^2 \Pi^{(0)}.\label{eq:be_ODE_stan}
\eeq
The solution to this ODE at arbitrary conformal time can be written in an explicit integral form
\begin{align}\label{eq:Qateta}
    &P_{\pm}^{(0)}(\eta,\bk,\hn)= -\frac32 (\bs{\epsilon}_{\pm}\cdot \hk)^2 \nonumber\\
    &\quad \times \int_0^{\eta} d \eta' \dot{\tau} ' e^{- \int_{\eta'}^{\eta} d \eta'' \dot{\tau}''}  e^{- i \bs{k} \cdot \hn (\eta - \eta')} \Pi^{(0)}(\bm{k}, \eta'),
\end{align}
where $\dot{\tau} \equiv \dot{\tau}^{(0)}$ is implied. The real-space solution today (at conformal time $\eta_0$), at the origin of coordinates, is then
\begin{align}
&P_{\pm}^{(0)}(\eta_0, \bs{x} = 0, \hn) = \int Dk ~P^{(0)}_{\pm}(\eta_0, \bk, \hn) \nonumber\\
&= -\frac32  \int Dk \int_0^{\eta_0}  d \eta ~g(\eta)~(\bs{\epsilon}_{\pm}\cdot \hk)^2 e^{- i \bs{k} \cdot \hn \chi} \Pi^{(0)}(\bm{k}, \eta),~~~~\label{eq:los-sol}
\end{align}
where from here on $\chi \equiv \eta_0 - \eta$ and $g(\eta)$ is the standard visibility function,  
\begin{align}
g(\eta) \equiv \dot{\tau}^{(0)}(\eta) \exp\left(- \int_\eta^{\eta_0}d \eta'~\dot{\tau}^{(0)}(\eta')\right).  
\end{align}
In Appendix \ref{app:spin}, we derive the following useful identity,
\begin{align}
-&(\hk \cdot \bs{\epsilon}_{\pm}(\hn))^2 e^{-i \bs{k} \cdot \hn\chi}  = \nonumber \\
&4\pi \sum_{\ell\geq 2, m} \frac{(-i)^{\ell}}{2} \sqrt{\frac{(\ell+2)!}{(\ell-2)!}} \frac{j_{\ell}(k \chi)}{(k \chi)^2} Y_{\ell m}^*(\hk) Y_{\ell m}^{\pm 2}(\hn) ,\label{eq:plane-wave-spin2}
\end{align} 
which is the spin-2 equivalent of the well-known scalar plane-wave expansion formula, 
\begin{align}
e^{-i \bs{k} \cdot \hn\chi} 
&= 4\pi \sum_{\ell m} (-i)^{\ell} j_{\ell}(k \chi) Y_{\ell m}^*(\hk) Y_{\ell m}(\hn).\label{eq:plane-wave}
\end{align} 
Inserting this result into Eq.~\eqref{eq:los-sol}, we can directly read off the coefficient of $P_{\pm}^{(0)}(\hn)$ on the spin-2 spherical harmonic basis, which is precisely $E_{\ell m}^{(0)} \pm i B_{\ell m}^{(0)}$. We thus obtain the well-known standard solution \cite{seljak96a},
\begin{align}\label{eq:E0}
    E^{(0)}_{\ell m} \pm i B_{\ell m}^{(0)} =4\pi(-i)^\ell  \int\!\!Dk ~\Delta_{E\ell}(k)~Y_{\ell m}^*(\hk)~\zeta({\bm k}),
\end{align}
where $\zeta(\bk)$ is the primordial curvature perturbation,
\begin{align}\label{eq:mathcalP}
\Delta_{E\ell}(k)&\equiv \int_0^{\eta_0} d\eta ~g(\eta)~\mathcal{E}_\ell(\eta,k),\\
        \mathcal{E}_\ell(\eta,k)&\equiv\frac{3}{4}\sqrt{\frac{(\ell+2)!}{(\ell-2)!}}\frac{j_{\ell}(k\chi)}{(k\chi)^2}\widetilde{\Pi}^{(0)}(\eta,k),
\end{align}
and $\widetilde{\Pi}^{(0)}$ is the transfer function of $\Pi^{(0)}$, defined through $\Pi^{(0)}(\bk) \equiv \widetilde{\Pi}^{(0)}(k) \zeta(\bk)$. Since the right-hand-side of Eq.~\eqref{eq:E0} is independent of the $\pm$ sign, we find that $B_{\ell m}^{(0)} = 0$, and Eq.~\eqref{eq:E0} in fact just gives $E_{\ell m}^{(0)}$.

The canonical power spectra are then (we subscript terms related to $\Theta$ with $T$)
\begin{align}
    \langle E_{\ell m}^{*(0)} E^{(0)}_{\ell' m'}\rangle&\equiv \delta_{\ell \ell'}\delta_{m m'}C_{EE\ell}^{(0)},\\
    \langle \Theta_{\ell m}^{*(0)} E^{(0)}_{\ell' m'}\rangle&\equiv \delta_{\ell \ell'}\delta_{m m'}C_{TE\ell}^{(0)},
\end{align}
with
\begin{align}
    C_{EE\ell}^{(0)}&=4\pi \int Dk ~\left[\Delta_{E\ell}^{(0)}(k)\right]^2P_{\zeta}(k),\\
    C_{TE\ell}^{(0)}&=4\pi \int Dk ~\Delta_{E\ell}^{(0)}(k)\Delta_{T\ell}^{(0)}(k)~P_{\zeta}(k),
\end{align}
where $\Theta^{(0)}_{\ell m}$ and $\Delta_{T\ell}^{(0)}$ are the standard temperature anisotropy and transfer function defined in Eq.~(33)~$\&$~(34) in Paper~II respectively (or cf. Ref.~\cite{seljak96a}).

\subsection{Polarization tensor due to perturbed recombination}
We now consider a more general free-electron fraction, $x_e(\bs{r})= x_e^{(0)}(1+\delta^{(1)}_e(\bs{r}))$, where $x_e^{(0)}$ is the standard, homogeneous free-electron fraction, and $\delta_e^{(1)}(\bs{r})$ is a general (inhomogeneous) perturbation, which may in general also have a non-vanishing homogeneous part $\langle \delta_e^{(1)} \rangle \neq 0$. 

We shall compute the polarization field at linear order in $\delta_e^{(1)}$, i.e.~write $P_{\pm} \approx P_{\pm}^{(0)} + P_{\pm}^{(1)}$, where $P_{\pm}^{(1)}$ is linear in $\delta_e^{(1)}$, and write a similar linear-expansion approximation for all other fields. At linear order in $\delta_e^{(1)}$, the polarization Boltzmann equation Eq.~\eqref{eq:CBE P_pm} becomes
\beq
\dot{P}_{\pm}^{(1)} + \hn \cdot \bs{\nabla} P_{\pm}^{(1)} + \dot{\tau}^{(0)} P_{\pm}^{(1)} = \dot{\tau}^{(0)} \left[S_{\pm}^{(1)d} + S_{\pm}^{(1)f}\right], \label{eq:be_ODE_pert}
\eeq
where we have split the source term on the right-hand-side into a ``direct" piece, directly proportional to the perturbed free-electron fraction $\delta_e^{(1)}$,
\barr
S_{\pm}^{(1)d}(\eta, \bs{x}, \hn) &\equiv&  \delta_e^{(1)}(\eta, \bs{x})  \mathcal{P}_{\pm}^{(0)}(\eta, \bs{x},\hn),~~ \label{eq:S1direct}
\earr
where we have defined
\barr
\mathcal{P}_{\pm}^{(0)}(\eta, \bx, \hn) &\equiv& \epsilon_{\pm}^i(\hn) \epsilon_{\pm}^j(\hn)\mathcal{P}_{ij}^{(0)}(\eta, \bs{x},\hn), \\
\mathcal{P}_{ij}^{(0)}(\eta, \bs{x},\hn) &\equiv& P_{ij}^{(0)}(\eta, \bs{x}, \hn) - \Pi_{ij}^{(0)}(\eta, \bs{x}), 
\earr
and a ``feedback" piece $S_{\pm}^{(1)f} \equiv -\epsilon_{\pm}^i \epsilon_{\pm}^j \Pi_{ij}^{(1)}$, still linear in $\delta_e^{(1)}$, but only indirectly through $\Pi_{ij}^{(1)}$.

The direct term depends on zero-th-order fields which can be readily extracted from codes like \texttt{CLASS} \citep{CLASS}. In contrast, computing the feedback term would require explicitly solving a Boltzmann hierarchy for the perturbed temperature and polarization fields. 
Like previous studies \cite{dvorkin13a, jensen23a}, we will only consider the direct term in this paper. Proceeding as in the standard case, we thus obtain the polarization helicities at the origin, today, from a line-of-sight integral
\begin{align}
    &P_{\pm}^{(1)}(\eta_0,{\bm x}=0,\hn)\approx\nonumber\\
    &\quad\quad\quad\int Dk\int_0^{\eta_0} d\eta ~g(\eta)~ S_{\pm}^{(1)d}(\eta, \bk, \hn) ~e^{-i{\bm k}\cdot\hn\chi}. \label{eq:P_pm^1d}
\end{align}
Note that for inhomogeneous $\delta_e^{(1)}$, the Fourier components of the source function $S_{\pm}^{(1)d}(\bk)$ are obtained from a convolution.

\section{Perturbed angular power spectra}\label{sec:powerspec}

In this section we compute the perturbations to CMB temperature and $E$-mode polarization power spectra at linear order in free-electron fraction perturbations, $C_{\ell}=C_{\ell}^{(0)}+C_{\ell}^{(1)}$. Note that, even though the $B$-mode polarization does not vanish at first order in $\delta_e^{(1)}$, its cross-correlation with $T$ and $E$ is guaranteed to vanish from pure symmetry consideration. The $B$-mode auto power spectrum $C_{BB \ell}^{(2)}$ is at least quadratic in $\delta_e^{(1)}$, thus parametrically suppressed, and we defer its calculation to Appendix~\ref{app:powerspec}. In this section, we thus focus specifically on the linear-order perturbations to $TE$ and $EE$ power spectra, which take the form
\begin{align}
    C_{TE\ell}^{(1)} \delta_{\ell \ell'}\delta_{m m'}&\equiv\langle\Theta_{\ell' m'}^{*(0)} E_{\ell m}^{(1)}\rangle+\langle\Theta_{\ell' m'}^{*(1)} E_{\ell m}^{(0)}\rangle, \label{eq:ClTE1}\\
    C_{EE\ell}^{(1)} \delta_{\ell \ell'}\delta_{m m'}&\equiv2\langle E_{\ell' m'}^{*(0)} E_{\ell m}^{(1)}\rangle,\label{eq:ClEE1}
\end{align}
where $\Theta_{\ell m}^{(1)}$ and $E_{\ell m}^{(1)}$ are the perturbed temperature and $E$-mode polarization fields at linear order in the free-electron perturbation. 

We start by considering the contributions $X_{\ell m}^{(1) \rm hom}$ and $C_{\ell}^{(1) \rm hom}$ to the perturbed $T$ and $E$ fields and their (cross) power spectra arising from the homogeneous part $\overline{\delta}_e$ of the free-electron fraction perturbation. This allows us to make contact with exact numerical results from  \texttt{CLASS}, which is capable of dealing with arbitrary homogeneous ionization histories, and check the accuracy of our approximations. We then discuss the contributions $X_{\ell m}^{(1) \rm inh}$ and $C_{\ell}^{(1) \rm inh}$ arising from the inhomogeneous part of free-electron fraction perturbations $\delta_e^{\rm inh}$. In particular, we will see that, even though $\delta_e^{\rm inh}$ is comparable in magnitude to $\overline{\delta}_e$, the perturbed power spectra $C_{\ell}^{(1)\rm inh}$ are significantly suppressed relative to their $C_\ell^{(1) \rm hom}$ counterparts.

\subsection{Homogeneous \texorpdfstring{$\delta_e$}{de} contributions}\label{sec:homo_de}

To quantify the approximation of neglecting the feedback term, and as a general crosscheck for our derivations, we first consider the special case where $\delta_e^{(1)}(\eta, \bs{x}) = \overline{\delta}_e^{(1)}(\eta)$ is homogeneous. In this case, we may compare the perturbed power spectra we obtain from our approximate perturbative approach with the exact (non-perturbative) output from \texttt{CLASS}. We emphasize, in particular, that even in the perturbative regime, \texttt{CLASS} effectively accounts for both the ``direct" and ``feedback" terms.

For a homogeneous $\delta_e^{(1)}$, the direct source term [Eq.~\eqref{eq:S1direct}] has the simple Fourier components
\barr
S_{\pm}^{(1)d, \rm hom}(\eta, \bk) &=&  \overline{\delta}_e^{(1)}(\eta)\mathcal{P}_{\pm}^{(0)}(\eta, \bk)\nonumber\\
\mathcal{P}_{\pm}^{(0)}(\eta, \bk) &\equiv & -\epsilon_{\pm}^i \epsilon_{\pm}^j \Pi_{ij}^{(0)}(\eta, \bk) - P_{\pm}^{(0)}(\eta, \bk), \label{eq:mathcalP0}
\earr
where the $\hn$ dependence is implicit. The first term in the right-hand side is given by Eq.~\eqref{eq:Pi_ij^0}, and the second term by Eq.~\eqref{eq:Qateta}, which we may rewrite in the form
\begin{align}
    &e^{-i{\bm k}\cdot\hn \chi}P_{\pm}^{(0)}(\eta, \bk)= -\frac32  (\bs{\epsilon}_{\pm} \cdot \hk)^2\nonumber\\
    &\quad\quad \times \frac{\dot{\tau}^{(0)}(\eta)}{g(\eta)} \int_0^{\eta} d \eta' g(\eta')  e^{- i \bs{k} \cdot \hn \chi'} \Pi^{(0)}(\bm{k}, \eta').
\end{align}
Moreover using the property that 
\beq
\frac{\dot{\tau}^{(0)}(\eta)}{g(\eta)}~ \int_0^\eta d\eta' g(\eta') = 1, 
\eeq
we obtain
\begin{align}
&e^{-i{\bm k}\cdot\hn \chi} \mathcal{P}_{\pm}^{(0)}(\eta, \bk, \hn) = \frac32 (\bs{\epsilon}_{\pm} \cdot \hk)^2 \frac{\dot{\tau}^{(0)}(\eta)}{g(\eta)}\int_0^{\eta} d \eta' g(\eta') \nonumber\\
&\quad\times \left[e^{- i \bs{k} \cdot \hn \chi'}\Pi^{(0)}(\bm{k}, \eta') - e^{- i \bs{k} \cdot \hn \chi}\Pi^{(0)}(\bm{k}, \eta)\right].~~
\end{align}
Using again the spin-2 plane-wave expansion [Eq.~\eqref{eq:plane-wave-spin2}], we may rewrite this as
\begin{align}
&e^{-i{\bm k}\cdot\hn \chi} \mathcal{P}_{\pm}^{(0)}(\eta, \bk, \hn) = \nonumber\\
&\quad\quad \quad\quad4 \pi \sum_{\ell m} (-i)^{\ell} \mathcal{J}_{E \ell}(\eta, k) Y_{\ell m}^*(\hk) Y_{\ell m}^{\pm2}(\hn)\zeta(\bk),~~~~~\label{eq:mathcalPO-harmonic}\\
&\mathcal{J}_{E\ell}(\eta,k)\equiv\frac{\dot{\tau}^{(0)}(\eta)}{g(\eta)}\int_0^\eta d\eta' g(\eta')\left(\mathcal{E}_\ell(\eta,k)-\mathcal{E}_\ell(\eta',k)\right) ,\label{eq:mathcalJ_E}
\end{align}
where $\mathcal{E}_\ell(\eta, k)$ was defined in Eq.~\eqref{eq:mathcalP}. Inserting this result into Eq.~\eqref{eq:P_pm^1d}, we may directly read off the coefficient of $Y_{\ell m}^{\pm 2}(\hn)$, which is $(E_{\ell m}^{(1) \rm hom} \pm i B_{\ell m}^{(1) \rm hom})$. As in the standard case, this coefficient is independent of the $\pm$ sign, from which we deduce that $B_{\ell m}^{(1) \rm hom} = 0$, as expected, and obtain
\begin{align}
E_{\ell m}^{(1) \rm hom}\approx&4\pi (-i)^\ell\int Dk~  
 \Delta_{E\ell}^{(1)\rm hom}(k) Y_{\ell m}^*(\hk)\zeta({\bm k}), \label{eq:Elm^1hom}\\
 \Delta_{E\ell}^{(1)\rm hom}(k)\equiv& \int_0^{\eta_0} d\eta~g(\eta) \overline{ \delta}_e^{(1)}(\eta)\mathcal{J}_{E\ell}(\eta,k).\label{eq:del_q_ell}
\end{align}
By switching the order of integration, one may also write $\Delta_{E\ell}^{(1)\rm hom}(k)$ in the equivalent form
\begin{align}
    \Delta_{E\ell}^{(1)\rm hom}(k)&= \int_0^{\eta_0} \!\!\!d\eta~g(\eta) \mathcal{E}_\ell(\eta,k)\left(\overline{\delta}_e^{(1)}(\eta)-\overline{\mathcal{D}}^{(1)}_e(\eta)\right),\\
\overline{\mathcal{D}}^{(1)}_e(\eta)&\equiv\int_{\eta}^{\eta_0} d \eta'~ \dot{\tau}^{(0)}(\eta') \overline{\delta}^{(1)}_e(\eta').
\end{align}
We compute $\Delta_{E\ell}^{(1)\rm hom}$ both ways and ensure they agree.

\begin{figure*}[htp]
\includegraphics[trim={0.7 0 0.4cm 0},width=\columnwidth]{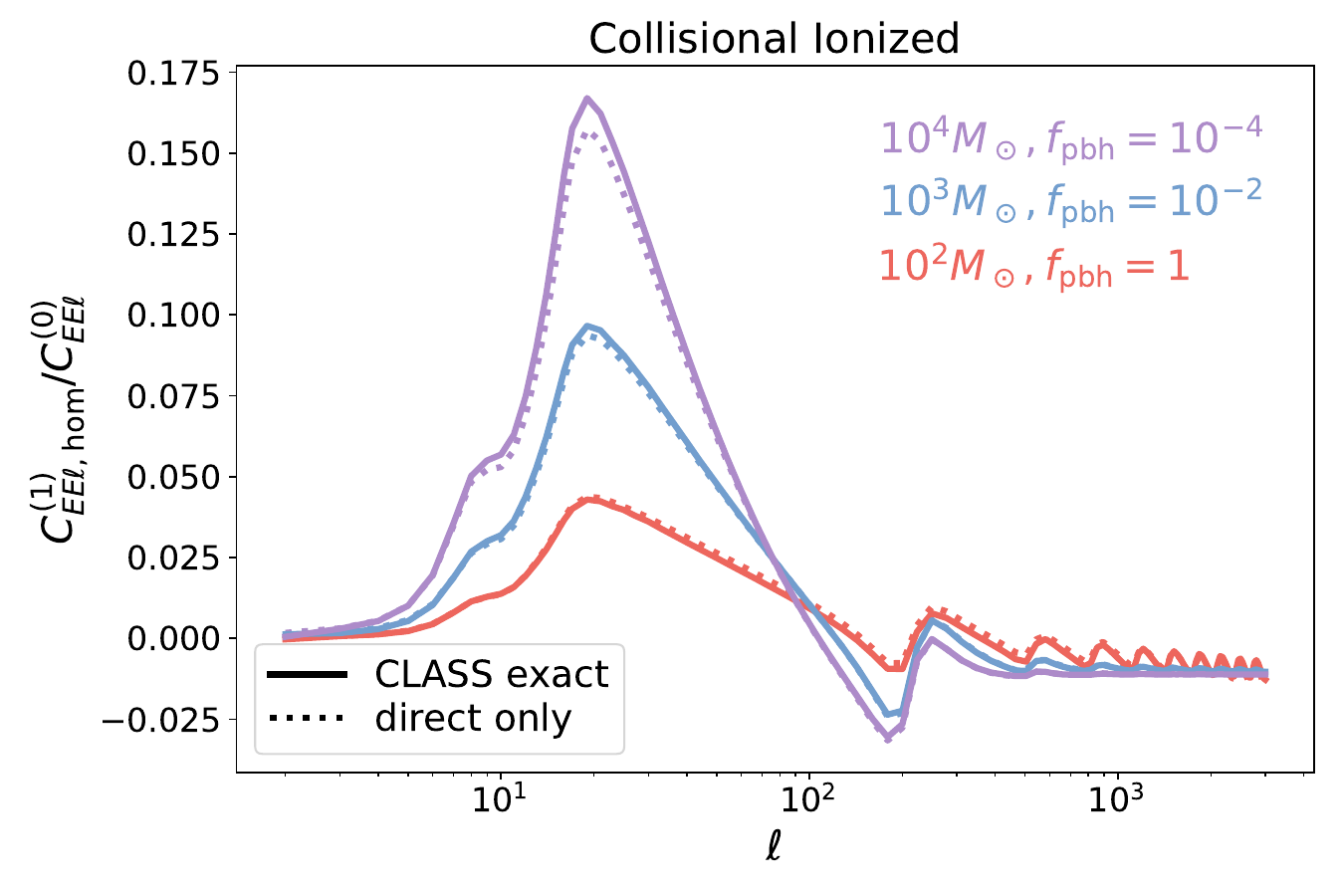}
\includegraphics[trim={0.7 0 0.4cm 0},width=\columnwidth]{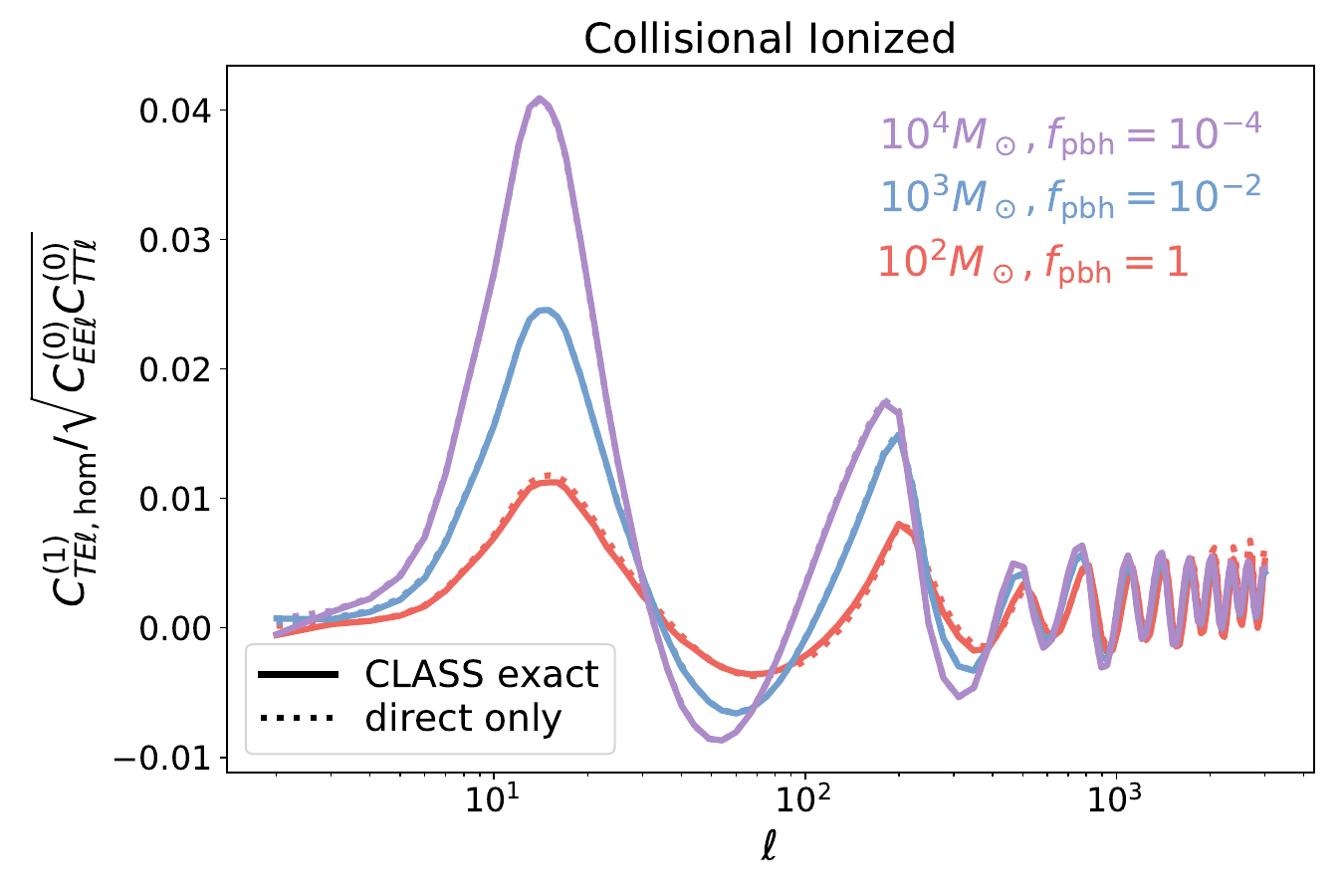}
\caption{\label{fig:class_v_comm} Normalized change to the $EE$ (left) and $TE$ (right) anisotropy power spectra resulting from the homogeneous part of the perturbation to the free-electron fraction, $\overline{\delta_e}(\eta)$, for various PBH masses and abundances. We compare the exact non-perturbative effect extracted from \texttt{CLASS} to the perturbative solution including only the ``direct" source term discussed in Sec.~\ref{sec:homo_de}. Our approximation of neglecting the ``feedback'' term is quite accurate and we assume this carries over for the inhomogeneous free-electron fraction case.}
\end{figure*}

We may now compute the approximate perturbations to the angular power spectra defined in Eqs.~\eqref{eq:ClTE1}-\eqref{eq:ClEE1}, with $E_{\ell m}^{(1) \rm hom}$ given in Eq.~\eqref{eq:Elm^1hom} above, and $\Theta^{(1)\rm hom}_{\ell m}$ taking a similar form, with the corresponding $\Delta_{T\ell}^{(1)\rm hom}$ and $\mathcal{J}_{T\ell}$ given in Eqs.~(53) and (54) of Paper II, respectively. We then find
\begin{align}
    &C_{TE\ell}^{(1)\rm hom}\approx 4 \pi \int Dk ~P_{\zeta}(k)\nonumber\\
    &\times \left(\Delta_{T \ell}^{(0)}(k) \Delta_{E\ell}^{(1) \rm hom}(k)+ \Delta_{T\ell}^{(1)\rm hom}(k) \Delta_{E\ell}^{(0)}(k)\right),\\
    &C_{EE\ell}^{(1)\rm hom} \approx 8 \pi \int Dk ~P_{\zeta}(k) \Delta_{E\ell}^{(0)}(k) \Delta_{E\ell}^{(1) \rm hom}(k).
\end{align}
We computed $C^{(1)\rm hom}_{TE \ell}$ and $C^{(1)\rm hom}_{EE \ell}$ using the homogeneous free-electron fraction induced by accreting PBHs, as calculated in AK17. As shown in Fig~\ref{fig:class_v_comm}, we compare these results to the exact, non-perturbative output from \texttt{CLASS}. We find very good agreement for a broad range of PBH masses, with accuracy on the $C_\ell^{(1)}$ better than $\sim$10$\%$. This indicates that both the linear-order expansion and the neglect of the ``feedback'' term are accurate approximations. Of course there is no guarantee that this also holds for an inhomogeneous free-electron perturbation, but this result does provide confidence that this approximation is a good first step, especially considering the large theoretical uncertainty in the PBH accretion model itself \cite{yacine17a}. 

\subsection{Inhomogeneous \texorpdfstring{$\delta_e$}{de} contributions}

The remainder of this section provides a detailed calculation of the perturbations to $TE$ and $EE$ power spectra resulting from the inhomogeneous part of the free-electron perturbation $\delta_e^{\rm inh}$. This calculation may be skipped by the reader primarily interested in the trispectrum calculations and results. The main outcome is that we find $C_{\ell}^{(1)\rm inh} \ll C_\ell^{(1)\rm hom}$, for $TE$ and $EE$ spectra, as we had already found for the temperature power spectrum in Paper II, despite the fact that $\delta_e^{\rm inh} \sim \overline{\delta}_e$. This implies that the inhomogeneous part of $\delta_e$ would have a negligible effect on CMB power spectra limits to PBHs, which ignore these contributions \cite{ricotti08a, yacine17a, poulin17a}. It is worthwhile pointing out an interesting geometric result: we prove that, in the limit that the free-electron fraction perturbation depends on the local magnitude of the relative velocity field (i.e.~if energy deposition is spatially on-the-spot), or of any longitudinal vector field in general, the two-point functions $\av{X^{(0)} E^{(1)\rm inh}}$ vanish identically, for $X^{(0)} = \Theta^{(0)}, E^{(0)}$. This implies that $C_{EE\ell}^{(1)\rm inh} = 0$ and that $C_{TE\ell}^{(1) \rm inh} = \av{\Theta_{\ell m}^{(1)\rm inh} E_{\ell m}^{(0)}}$ in this limit.

\subsubsection{Universal dependence of linear-order power spectra on PBH luminosity}

In this subsection we work in real space, to derive general geometric properties of cross-correlation functions, in a similar spirit as in Ref.~\cite{YAH_24}. 

The unperturbed temperature and $E$-mode polarization fields are both linear in the primordial curvature perturbation $\bs{\zeta}$ and thus take the form,
\beq
X^{(0)}(\eta_0, \bs{0}, \hn) = \int d^3 \bs{x} ~ \Delta_X^{(0)}(\bs{x}, \hn) \zeta(\bs{x}),
\eeq
where $X\in\{\Theta,E\}$, and $\Delta_X(\bs{x}, \hn)$ is the relevant real-space transfer function. 

In real space, Eq.~\eqref{eq:P_pm^1d} takes the form
\barr
P_{\pm}^{(1)}(\eta_0,\bs{0},\hn)\approx \int_0^{\eta_0} d\eta ~g(\eta)~ S_{\pm}^{(1)d}(\eta, \bs{x} = -\chi \hn, \hn), \label{eq:P_pm^1d-real}
\earr
where the source function $S_{\pm}^{(1)d}$ is given in Eq.~\eqref{eq:S1direct}. This source function is proportional to $\delta_e^{(1)} \mathcal{P}_{ij}$ evaluated at the same spatial position. 

In the context of accreting PBHs as was discussed in Section~\ref{sec:paperI}, in the limit that perturbations to the ionization history are linear in the energy injection rate, we can write $\delta_e$ generally in real space,
\beq
\delta_e(\eta, \bs{x}) = \int_0^{\eta} d \eta' \int d^3 x' ~ \mathcal{G}_e(\eta, \eta', |\bs{x}' - \bs{x}|) L(v(\eta', \bs{x}')), 
\eeq
where $\mathcal{G}_e(\eta, \eta', r)$ is a general Green's function response and $L$ is the PBH luminosity, which depends on the magnitude $v$ of the local velocity $\bs{v}$ of baryons relative to PBHs. Therefore, we see that the cross-correlations $\langle X^{(0)}(\hn) P_{\pm}^{(1)}(\hn')\rangle$, or equivalently $\langle X^{(0)}_{\ell m} E^{(1)}_{\ell' m'}\rangle$, all depend on the following integral,
\barr
\Psi_{\pm}(\bs{x}_1, \bs{x}_2, \hn) \equiv \epsilon_{\pm}^i \epsilon_{\pm}^j \int d^3 x_3 \mathcal{G}_e(|\bs{x}_3 - \bs{x}_2|)\nonumber\\
 \times \Xi_{ij}(\bs{x}_1, \bs{x}_2, \bs{x}_3, \hn), \label{eq:Psi_pm}
\earr
where $\Xi_{ij}$ is the ``fundamental" 3-point correlation function,
\beq
\Xi_{ij}(\bs{x}_1, \bs{x}_2, \bs{x}_3, \hn) \equiv \av{\zeta(\bs{x}_1) \mathcal{P}_{ij}(\bs{x}_2, \hn) L(v(\bs{x}_3)},
\eeq
where the $(0)$ superscript on $\mathcal{P}_{ij}$ is omitted for conciseness. Note that we do not explicitly write the time dependence, as it is not relevant to our argument.

To compute $\Xi_{ij}$, we will first compute the constrained 2-point correlation of $\zeta(\bs{x}_1)$ and $\mathcal{P}_{ij}(\bs{x}_2)$ under the constraint that $\bs{v}(\bs{x}_3)$ takes a fixed value $\bs{v}$, which we denote by $\av{\zeta(\bs{x}_1) \mathcal{P}_{ij}(\bs{x}_2)|\bs{v}(\bs{x}_3) = \bs{v}}$. We will then multiply the result by $L(v^2)$ and average it over the Gaussian distribution of $\bs{v}$. We start by defining the following variables:
\barr
\zeta' &\equiv& \zeta(\bs{x}_1) - \frac{\av{\zeta(\bs{x}_1) \bs{v}(\bs{x}_3)}}{\frac13 \av{v^2}}\cdot \bs{v}(\bs{x}_3), \\
\mathcal{P}_{ij}' &\equiv& \mathcal{P}_{ij}(\bs{x}_2) - \frac{\av{\mathcal{P}_{ij}(\bs{x}_2) \bs{v}(\bs{x}_3)}}{\frac13 \av{v^2}}\cdot \bs{v}(\bs{x}_3).
\earr
By construction, $\zeta'$ and $\mathcal{P}_{ij}'$ are uncorrelated with $\bs{v}(\bs{x}_3)$. This implies that their constrained averages and correlation function are, in fact, independent of $\bs{v}$, and can thus be obtained by averaging over the distribution of $\bs{v}(\bs{x}_3)$. After some algebra, we obtain
\begin{align}
\av{\zeta(\bs{x}_1) \mathcal{P}_{ij}(\bs{x}_2)|\bs{v}(\bs{x}_3) = \bs{v}} = \av{\zeta(\bs{x}_1) \mathcal{P}_{ij}(\bs{x}_2)} \nonumber\\
+ \frac{\av{\zeta(\bs{x}_1) v_k(\bs{x}_3)}\av{\mathcal{P}_{ij}(\bs{x}_2) v_l(\bs{x}_3)}}{\left(\frac13 \av{v^2}\right)^2}\left(v_k v_l - \frac13 \delta_{kl} \av{v^2} \right).
\end{align}
We now multiply this result by $L(v)$ and integrate over the isotropic Gaussian distribution of $\bs{v}$. Since $L(v)$ only depends on the magnitude of $\bs{v}$, we arrive at
\barr
\Xi_{ij}(\bs{x}_1, \bs{x}_2, \bs{x}_3, \hn) = \av{\zeta(\bs{x}_1) \mathcal{P}_{ij}(\bs{x}_2)} \av{L} \nonumber\\
+ \frac{\av{\zeta(\bs{x}_1) \bs{v}(\bs{x}_3)}\cdot\av{\mathcal{P}_{ij}(\bs{x}_2) \bs{v}(\bs{x}_3)}}{\frac13 \av{v^2}} L_2, \label{eq:Xi_ij}
\earr
where $L_2$ is the second moment of the PBH luminosity,
\beq
L_2 \equiv \frac{\av{v^2 L(v)}}{\av{v^2}} - \av{L(v)}.
\eeq
We have therefore arrived at a powerful result: the ``fundamental" 3-point function $\Xi_{ij}$, hence $\av{\Theta^{(0)} E^{(1)}}$ and $\av{E^{(0)} E^{(1)}}$, only depend on the PBH luminosity through its average over relative velocities $\av{L}$ and its second moment $L_2$. In fact the above reasoning would hold equally for the perturbed temperature field and the correlations $\av{\Theta^{(0)} \Theta^{(1)}}$ and $\av{E^{(0)} \Theta^{(1)}}$. As a consequence, the linear-order perturbations to all CMB $T$ and $E$ power spectra only depend on $L(v)$ through $\av{L}$ and $L_2$. Therefore, one may approximate $L(v)$ by a simple quadratic function in $v$ with the correct average and second moment, and obtain the \emph{exact} CMB power spectra at linear-order in free-electron perturbations. Of course this result no longer holds for trispectra, for which a similar reasoning would show that they depend on $L_2$ and on the fourth velocity moment of $L$. 

\subsubsection{\texorpdfstring{$C_{TE \ell}^{(1)}$}{CTE} and \texorpdfstring{$C_{EE \ell}^{(1)}$}{CEE} from inhomogeneities in \texorpdfstring{$\delta_e$}{de}}\label{subsec:CTE}

The first term in Eq.~\eqref{eq:Xi_ij} corresponds to the effect of the homogeneous part of $\delta_e$, sourced by the averaged PBH luminosity $\av{L}$, which we already addressed in the previous section. The second term in Eq.~\eqref{eq:Xi_ij} corresponds to the effect of the inhomogeneous part of $\delta_e$, sourced by the non-trivial velocity-dependence of $L$, implying a non-vanishing $L_2$. This term is proportional to the two-point correlation function of $\mathcal{P}_{ij}(\hn)$ and $\bs{v}$. The former tensor is symmetric and trace-free, and the latter vector field is longitudinal (i.e.~curl-free). By explicitly writing the 2-point correlation function as the Fourier transform of the cross-power spectrum, one can see that it must take the geometric form
\barr
\av{\mathcal{P}_{ij}(\bs{x}_2, \hn) \bs{v}(\bs{x}_3)} = (\alpha_0 \bs{r} + \beta_0 \hn) (\hn_i \hn_j - \frac13 \delta_{ij})\nonumber\\
+(\alpha_1 \bs{r} + \beta_1 \hn) (r_i \hn_j + \hn_i r_j - \frac13 (\bs{r} \cdot \hn) \delta_{ij})\nonumber\\
+ (\alpha_2 \bs{r} + \beta_2 \hn) (r_i r_j - \frac13 r^2 \delta_{ij}),
\earr
where $\bs{r} \equiv \bs{x}_3 - \bs{x}_2$, and the coefficients $\alpha_\ell, \beta_\ell$ are all functions of $r$ and $\bs{r} \cdot \hn$ alone, finite at $\bs{r} \rightarrow 0$. Upon dotting this correlation onto  $\epsilon_{\pm}^i(\hn)\epsilon_{\pm}^j(\hn)$, as eventually required to compute power spectra, only the third term survives: 
\beq
\epsilon_{\pm}^i(\hn)\epsilon_{\pm}^j(\hn)\av{\mathcal{P}_{ij}(\bs{x}_2, \hn) \bs{v}(\bs{x}_3)} = (\alpha_2 \bs{r} + \beta_2 \hn) (\epsilon_{\pm}(\hn) \cdot \bs{r})^2.
\eeq
This result implies that this 2-point correlation function scales at least quadratically with $r$ at small separations, and in particular vanishes at zero separation ($\bs{x}_3 = \bs{x}_2)$. 

Since $\av{\Theta^{(0)} E^{(1)}}$ and $\av{E^{(0)} E^{(1)}}$ both depend on the integral \eqref{eq:Psi_pm}, which weighs the above 2-point correlation function by the Green's function $G_e(|\bs{x}_3 - \bx_2|)$, we reach the following important conclusion: in the limit of spatially on-the-spot ionization by the photons injected by accreting PBHs, i.e.~if $\mathcal{G}_e(r) \propto \delta_{\rm D}(r)$, the inhomogeneous part of $\delta_e$ does not contribute to $\av{\Theta^{(0)} E^{(1)}}$ and $\av{E^{(0)} E^{(1)}}$. This also applies to our factorized approximation discussed in Section~\ref{sec:paperI}, which can be seen as an on-the-spot ionization quadratic in the transverse vector field $\bs{u}$ (see Eq.~\eqref{eq:delta_e-u^2} and discussion around it).

As a consequence, in this limit, $C_{EE \ell}^{(1)} = C_{EE \ell}^{(1)\rm hom}$ computed earlier, and $C_{TE \ell}^{(1)}$ gets an additional contribution from $\delta_e^{\rm inh} \equiv \delta_e - \av{\delta_e}$ given by $C_{TE \ell}^{(1)\rm inh} = \av{E_{\ell m}^{(0)*} \Theta_{\ell m}^{(1) \rm inh}}$. This power spectrum is identical to $C_{TT\ell}^{(1)\rm inh}$ given in Eq.~(75) in Paper~II, with the substitution $\Delta_{T\ell}^{(0)} \rightarrow \Delta_{E \ell}^{(0)}$, that is,
\begin{align}
  C_{TE\ell}^{(1)\rm inh} &= - \frac{8 \pi}3 \fbh \int_0^{\eta} d\eta ~g(\eta) \gamma(\eta) \xi_\ell(\eta) , \label{eq:C1_inh}
\end{align}
where we have defined
\begin{align}
 \gamma(\eta) &\equiv \int Dk ~ P_\zeta(k) \Delta_e(\eta, k) \widetilde{v}_{b \gamma}(\eta, k)\label{eq:gamma}, \\
 \xi_\ell(\eta) &\equiv \int Dk ~P_\zeta(k) \Delta_e(\eta, k) \Delta_{E\ell}^{(0)}(k)  j_\ell'(k \chi),
\end{align}
and $\bs{v}_{b \gamma}(\bk) \equiv (\bs{v}_b - \bs{v}_\gamma)(\bk) = -i \hk \widetilde{v}_{b \gamma}(k) \zeta(\bk)$ is the baryon-photon slip term (c.f. Paper~II Sec. IV B).

\begin{figure}[ht]
\includegraphics[trim={0 0.5cm 0 0.5cm},width=\columnwidth]{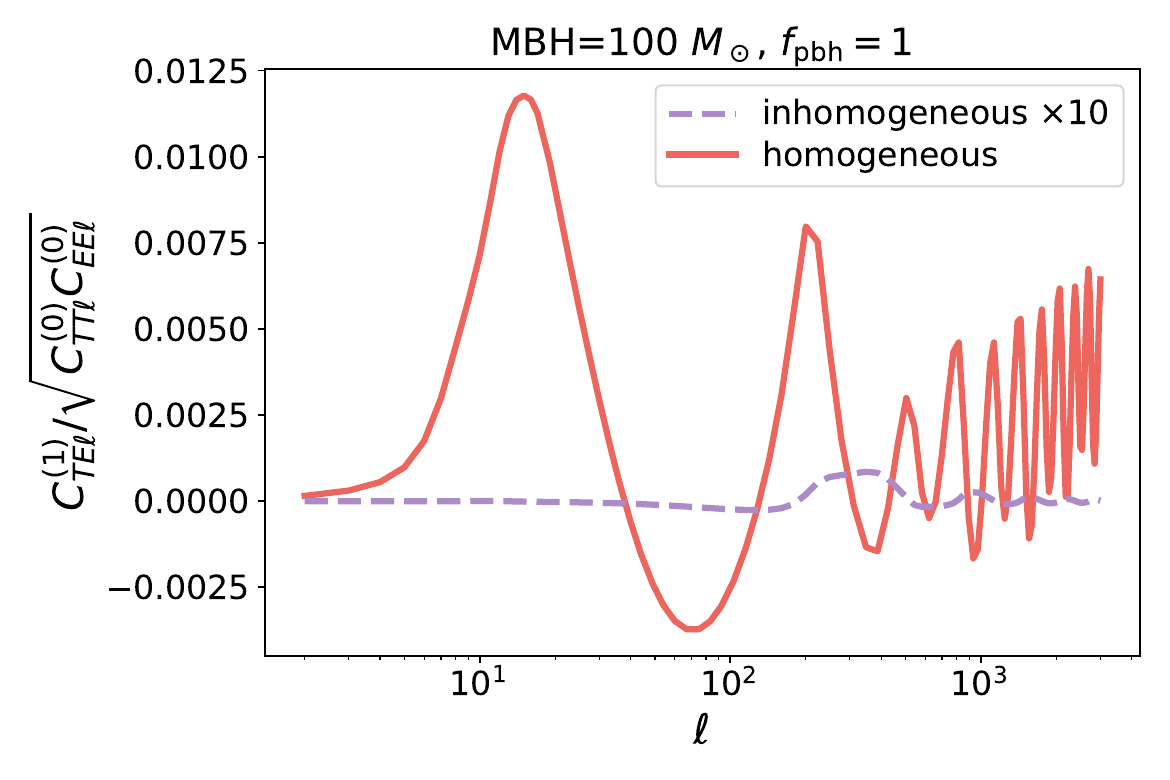}
\caption{\label{fig:AK17_jen} Fractional change to the $TE$ anisotropy power spectrum due to accreting PBHs of 100 $M_\odot$ comprising all the dark matter. We compare the contribution due to the inhomogeneous part of the ionization fraction perturbations, $C_{TE\ell}^{(1) \rm inh}$ (multiplied by a factor of 10 for clarity), calculated for the first time in this work, with the one arising from the homogeneous part of the free-electron fraction, $C_{TE\ell}^{(1)\rm hom}$, previously computed in AK17. The total effect on the power spectrum is a sum of both contributions. We see that contribution from inhomogeneities in recombination is negligible, implying that it would not appreciably change current 2-point constraints on accreting PBHs.}
\end{figure}

We computed $C_{TE\ell}^{(1)\rm inh}$ and show the result in Fig.~\ref{fig:AK17_jen}, where we compare this term to its counterpart $C_{TE\ell}^{(1)\rm hom}$ sourced by the homogeneous part of the free-electron fraction perturbation, for 100-$M_{\odot}$ accreting PBHs. We find that $C_{TE\ell}^{(1)\rm inh}$ is smaller by a factor of $\sim 10- 100$, depending on scale, relative to $C_{TE\ell}^{(1)\rm hom}$ for all black hole masses. Thus incorporating the inhomogeneous perturbation computed here will not affect current reported constraints on accreting PBHs at the level of 2-point statistics. 

\section{Temperature and polarization trispectra due to accreting PBHs}\label{sec:trispec}

We now proceed to the main part of this paper: the computation of CMB anisotropy trispectra sourced by inhomogeneously accreting PBHs. We follow closely the procedure used in Paper~II, generalizing quantities to incorporate both temperature and polarization, including both $E$ and $B$ modes, which are sourced by $\delta_e^{\rm inh}$. 

\subsection{E and B transfer functions}

\subsubsection{Definitions}
When neglecting nonlinearities such as lensing, the standard temperature and $E$-mode anisotropy is linearly related to the primordial curvature perturbation, through (cf. Eq~\eqref{eq:E0} and Paper~II),
\begin{align}
    X_{\ell m}^{(0)} &= \int Dk ~ T^{(0)}_{X\ell m}(\bk) ~ \zeta(\bk), \label{eq:X0_lm}\\
    T_{X\ell m}^{(0)}(\bk) &\equiv 4 \pi (-i)^{\ell} \Delta^{(0)}_{X\ell}(k) Y_{\ell m}^*(\hat{k}), \label{eq:TX0_lm}
\end{align}
where $X = T, E$.

We approximate the free-electron fraction perturbation due to accreting PBHs as quadratic in the primordial curvature perturbation as shown in Sec.~\ref{sec:paperI}. This implies the corresponding first order perturbation to polarization anisotropy will be \textit{cubic}. The goal of this section is to derive the cubic transfer functions,
$T_{X\ell m}^{(1)}(\bk_1, \bk_2, \bk_3)$, defined through
\begin{align}\label{eq:EB_transf_int}
    X_{\ell m}^{(1)\rm inh} = \fbh \int D(k_1 k_2 k_3) T_{X\ell m}^{(1)}(\bk_1, \bk_2, \bk_3)\nonumber\\
    \times \zeta(\bk_1) \zeta(\bk_2) \zeta(\bk_3),
\end{align}
for $X = E, B$, where the label ``inh" indicates that here we focus on the inhomogeneous-$\delta_e$ contribution to $E^{(1)}$ and $B^{(1)}$. Note that the corresponding function for $X = T$ was already derived in Paper II. 

As we saw in Paper II for $X = T$, and as will see for $X = E, B$, these cubic transfer functions can be recast in the form 
\begin{align}\label{eq:T_multipoles}
T_{X\ell m}^{(1)}(\bk_1, \bk_2, \bk_3) &= (4 \pi)^3 \sum_{\ell_1 \ell_2 \ell_3}(-i)^{\ell_1 + \ell_2 + \ell_3}\nonumber\\
&\times \sum_{m_1 m_2 m_3} T_{X\ell_1 \ell_2\ell_3; \ell}^{m_1 m_2 m_3; m}(k_1, k_2, k_3)\nonumber\\
&\times Y_{\ell_1 m_1}(\hat{k}_1)  Y_{\ell_2 m_2}(\hat{k}_2) Y_{\ell_3 m_3}(\hat{k}_3).
\end{align}

\subsubsection{Calculations}

We start by rewriting the direct source term \eqref{eq:S1direct} as a convolution between $\delta_e^{(1)\rm inh}$ and $\mathcal{P}_{\pm}^{(0)}$. Using Eq.~\eqref{eq:Te-def}, we arrive at the following triple integral: 
\begin{align}
    S_{\pm}^{(1)d}(\eta, \bk, \hat{n}) = \fbh \int D(k_1 k_2 k_3) \sld(\bk_1 + \bk_2 +\bk_3 - \bk) \nonumber\\
    \,\,\times T_e(\eta, \bk_1, \bk_2) \widetilde{\mathcal{P}}^{(0)}_{\pm}(\eta,\bk_3,\hn)\zeta(\bk_1) \zeta(\bk_2) \zeta(\bk_3),&
\end{align}
where we recall that tilded quantities denote standard transfer functions, i.e.~ratios of fields to the primordial curvature perturbation in Fourier space. 

Substituting $T_e(\eta, \bk_1, \bk_2)$ with Eq.~\eqref{eq:dele_pbh} (and recalling that this expression only holds for $\bk_1 + \bk_2 \neq 0$, and that $T_e(\bk_1, - \bk_2)$ vanishes), and inserting this expression into Eq.~\eqref{eq:P_pm^1d} and taking its spin-2 spherical-harmonic transform as in Eq.~\eqref{eq:E&B_decomp}, we may rewrite the fields $P_{\ell m}^{\pm(1)\rm inh} \equiv E_{\ell m}^{(1)\rm inh} \pm i B_{\ell m}^{(1)\rm inh}$ in the form of Eq.~\eqref{eq:EB_transf_int}, with
\begin{align}
    T_{P^{\pm}\ell m}^{(1)}(\bk_1, \bk_2, \bk_3) = \int_0^{\eta_0} d \eta ~ g(\eta)  \nonumber\\
    \times (\hk_1 \cdot \hk_2) \Delta_e(\eta,k_1)\Delta_e(\eta, k_2)\nonumber\\
    \times \int d^2 \hn ~ Y^{\pm 2*}_{\ell m}(\hn)  e^{-i\chi\hat{n}\cdot(\bk_1 + \bk_2 + \bk_3)} \widetilde{\mathcal{P}}^{(0)}_{\pm}(\eta, \bk_3,\hn).\label{eq:TPpm1_lm}
\end{align}
This expression only holds for $\bk_1 + \bk_2 \neq 0$; in addition, we have $T_{P^{\pm}\ell m}^{(1)}(\bk_1, -\bk_1, \bk_3) = 0$. Note that $T_{P^{\pm}\ell m}^{(1)}$ is symmetric under exchange of $\bk_1$ and $\bk_2$. 

We can now compute the harmonic coefficients of Eq.~\eqref{eq:TPpm1_lm} in the form Eq.~\eqref{eq:T_multipoles}. First, as in Paper~II, we rewrite (denoting $\bs{\chi} \equiv \chi \hat{n}$),
\begin{align}
(\bk_1 \cdot \bk_2) e^{-i \bs{\chi} \cdot (\bk_1 + \bk_2)} &= -\left[\bs{\nabla}_{\bs{\chi}}  e^{-i \bs{\chi} \cdot \bk_1}\right] \cdot  \left[\bs{\nabla}_{\bs{\chi}}  e^{-i \bs{\chi} \cdot \bk_2}\right] \nonumber\\
&=  -\partial_\chi\left(e^{-i \bs{\chi} \cdot \bk_1}\right) \partial_\chi \left(e^{-i \bs{\chi} \cdot \bk_2}\right)\nonumber\\
& \quad -  \frac1{\chi^2} \left[\bs{\nabla}_{\hat{n}} e^{-i \bs{\chi} \cdot \bk_1}\right] \cdot  \left[\bs{\nabla}_{\hat{n}}  e^{-i \bs{\chi} \cdot \bk_2}\right],
\end{align}
where $\bs{\nabla}_{\bs{\chi}}$ is the gradient with respect to $\bs{\chi}$. In the second equality, we split the gradient into its radial part $\hat{n} \partial_\chi$ and its angular part $\frac1{\chi} \bs{\nabla}_{\hat{n}}$. Then, using the Rayleigh expansion Eq.~\eqref{eq:plane-wave}, we have
\begin{align}
(\hk_1 \cdot \hk_2 ) e^{-i \chi \hn \cdot (\bk_1 + \bk_2)} = -(4 \pi)^2 \sum_{\ell_1 \ell_2} (-i)^{\ell_1 + \ell_2} \nonumber\\
\sum_{m_1 m_2}Y_{\ell_1 m_1}(\hk_1) Y_{\ell_2 m_2}(\hk_2)\nonumber\\
\times \Big{[}j_{\ell_1}'(\chi k_1) j_{\ell_2}'(\chi k_2) Y_{\ell_1 m_1}^*(\hn) Y_{\ell_2 m_2}^*(\hn)\nonumber\\
+ \frac{j_{\ell_1}(\chi k_1)}{\chi k_1} \frac{j_{\ell_1}(\chi k_2)}{\chi k_2} \bs{\nabla}_{\hn} Y_{\ell_1 m_1}^*(\hn)  \cdot \bs{\nabla}_{\hn} Y_{\ell_2 m_2}^*(\hn) \Big{]}.
\end{align}
Using Eq.~\eqref{eq:mathcalPO-harmonic}, and the spherical harmonics property $Y_{\ell m}^{s*} = (-1)^{m + s}Y_{\ell -m}^{-s}$, we may rewrite Eq.~\eqref{eq:TPpm1_lm} in the form \eqref{eq:T_multipoles}, with
\barr
T_{P^{\pm} \ell_1 \ell_2, \ell_3 \ell_4}^{m_1 m_2 m_3 m_4} &= A_{\ell_1 \ell_2, \ell_3}(k_1, k_2, k_3) Q_{P^{\pm}\ell_1 \ell_2, \ell_3 \ell_4}^{m_1 m_2, m_3 m_4} \nonumber\\
   &+ \widetilde{A}_{\ell_1 \ell_2, \ell_3}(k_1, k_2, k_3) \widetilde{Q}_{P^{\pm}\ell_1 \ell_2, \ell_3 \ell_4}^{m_1 m_2, m_3 m_4}, \label{eq:Tmult-final-Ppm}
\earr
where the rotationally-invariant coefficients $A_{\ell_1 \ell_2, \ell_3}$ and $\widetilde{A}_{\ell_1 \ell_2, \ell_3}$ are given by
\begin{align}
&A_{\ell_1 \ell_2, \ell_3}(k_1, k_2, k_3) \equiv- \int\!\! d \eta ~ g(\eta) j_{\ell_1}'(\chi k_1)\Delta_e (\eta, k_1)\nonumber\\
&~~~\times j_{\ell_2}'(\chi k_2)\Delta_e(\eta, k_2) \mathcal{J}_{E\ell_3}(\eta,k_3),~~~~~\label{eq:A} \\
&\widetilde{A}_{\ell_1 \ell_2, \ell_3}(k_1, k_2, k_3) \equiv-  \int \!\!d \eta ~ g(\eta) ~f_{\ell_1} \frac{j_{\ell_1}(\chi k_1)}{\chi k_1}\Delta_e (\eta, k_1)\nonumber\\
&~~~\times f_{\ell_2} \frac{j_{\ell_2}(\chi k_2)}{\chi k_2} \Delta_e(\eta, k_2) \mathcal{J}_{E\ell_3}(\eta,k_3),~~~~~\label{eq:B}
\end{align}
where $f_{\ell} \equiv \sqrt{\ell (\ell +1)}$, and the purely geometric terms $Q_{P^{\pm}\ell_1 \ell_2, \ell_3, \ell_4}^{m_1 m_2, m_3, m_4}$ and $\widetilde{Q}_{P^{\pm}\ell_1 \ell_2, \ell_3, \ell_4}^{m_1 m_2, m_3, m_4}$ are integrals of the product of four spherical harmonics or their gradients:
\begin{align}
&Q_{P^{\pm}\ell_1 \ell_2, \ell_3\ell_4}^{m_1 m_2, m_3 m_4} \equiv \nonumber\\
&\int\! d^2 \hat{n} ~Y^{*}_{\ell_1 m_1}(\hat{n}) Y^*_{\ell_2 m_2}(\hat{n}) Y^{\mp 2*}_{\ell_3 m_3}(\hat{n}) Y_{\ell_4 m_4}^{\pm2*}(\hat{n}), \label{eq:Q_sym_P}\\
&\widetilde{Q}_{P^{\pm}\ell_1 \ell_2, \ell_3 \ell_4}^{m_1 m_2, m_3 m_4} \equiv \frac1{f_{\ell_1} f_{\ell_2}}\nonumber\\
&\times \int d^2 \hat{n} ~\bs{\nabla}_{\hat{n}}Y^*_{\ell_1 m_1}(\hat{n})\cdot \bs{\nabla}_{\hat{n}} Y^*_{\ell_2 m_2}(\hat{n}) Y^{\mp 2*}_{\ell_3 m_3}(\hat{n}) Y_{\ell_4 m_4}^{\pm2*}(\hat{n})\label{eq:Qt_sym_P}.
\end{align}
Note that we have included a factor of $f_{\ell_1} f_{\ell_2}$ in the $\widetilde{A}_{\ell_1 \ell_2 \ell_3}$ coefficient and divided the $\widetilde{Q}$ symbol by the same factor\footnote{Our notation is thus slightly different from Paper II.}, such that $\widetilde{Q}$ and $Q$ symbols are of the same order of magnitude, given that \citep{smith15a},
\begin{align}
\bs{\nabla}_{\hat{n}}Y_{\ell_1 m_1} \cdot \bs{\nabla}_{\hat{n}} Y_{\ell_2 m_2} &= -\frac12 f_{\ell_1} f_{\ell_2} \sum_{s = \pm1} Y_{\ell_1 m_1}^{s} ~Y_{\ell_2 m_2}^{-s}.
\end{align}

The $E$ and $B$ components of the polarization are then obtained from $E_{\ell m} = (P^+_{\ell m} + P^-_{\ell m})/2$ and $B_{\ell m} = (P^+_{\ell m} - P^-_{\ell m})/2i$, thus the harmonic coefficients of their perturbed transfer functions $T^{(1)}_{E/B}$ both take the form ($T^{(1)}_T$ is defined in Paper~II as Eq.~64)
\begin{align}
   T_{E/B\ell_1 \ell_2 \ell_3 \ell_4}^{m_1 m_2 m_3 m_4}&(k_1, k_2, k_3) =\nonumber\\
   &A_{\ell_1 \ell_2, \ell_3}(k_1, k_2, k_3) Q_{E/B\ell_1 \ell_2, \ell_3, \ell_4}^{m_1 m_2, m_3, m_4}\nonumber\\
   &+ \widetilde{A}_{\ell_1 \ell_2, \ell_3}(k_1, k_2, k_3) \widetilde{Q}_{E/B\ell_1 \ell_2, \ell_3, \ell_4}^{m_1 m_2, m_3, m_4}, \label{eq:Tmult-final}
\end{align}
where the geometric symbols are given explicitly by
\begin{align}
&Q_{E\ell_1 \ell_2, \ell_3\ell_4}^{m_1 m_2, m_3 m_4} \!\equiv\frac{1}{2}\int\! d^2 \hat{n} ~Y^{*}_{\ell_1 m_1}(\hat{n}) Y^*_{\ell_2 m_2}(\hat{n})\nonumber\\
&\quad\quad\times \left(Y^{- 2*}_{\ell_3 m_3}(\hat{n}) Y_{\ell_4 m_4}^{+2*}(\hat{n})+Y^{ +2*}_{\ell_3 m_3}(\hat{n}) Y_{\ell_4 m_4}^{-2*}(\hat{n})\right), \label{eq:Q_sym_E}\\
&\widetilde{Q}_{E\ell_1 \ell_2, \ell_3 \ell_4}^{m_1 m_2, m_3 m_4} \equiv -\frac{1}{2}\int d^2 \hat{n} ~\sum_{s = \pm 1} Y_{\ell_1 m_1}^{s*} Y_{\ell_2 m_2}^{-s}\nonumber\\
&\quad\quad\times \left(Y^{- 2*}_{\ell_3 m_3}(\hat{n}) Y_{\ell_4 m_4}^{+2*}(\hat{n})+Y^{ +2*}_{\ell_3 m_3}(\hat{n}) Y_{\ell_4 m_4}^{-2*}(\hat{n})\right),\label{eq:Qt_sym_E}\\
&Q_{B\ell_1 \ell_2, \ell_3;\ell_4}^{m_1 m_2, m_3; m_4} \!\equiv-\frac{i}{2}\int\! d^2 \hat{n} ~Y^{*}_{\ell_1 m_1}(\hat{n}) Y^*_{\ell_2 m_2}(\hat{n})\nonumber\\
&\quad\quad\times \left(Y^{- 2*}_{\ell_3 m_3}(\hat{n}) Y_{\ell_4 m_4}^{+2*}(\hat{n})-Y^{ +2*}_{\ell_3 m_3}(\hat{n}) Y_{\ell_4 m_4}^{-2*}(\hat{n})\right), \label{eq:Q_sym_B}\\
&\widetilde{Q}_{B\ell_1 \ell_2, \ell_3; \ell_4}^{m_1 m_2, m_3; m_4} \equiv \frac{i}{4}\int d^2 \hat{n} ~\sum_{s = \pm 1} Y_{\ell_1 m_1}^{s*} Y_{\ell_2 m_2}^{-s*}\nonumber\\
&\quad\quad\times \left(Y^{- 2*}_{\ell_3 m_3}(\hat{n}) Y_{\ell_4 m_4}^{+2*}(\hat{n})-Y^{ +2*}_{\ell_3 m_3}(\hat{n}) Y_{\ell_4 m_4}^{-2*}(\hat{n})\right).\label{eq:Qt_sym_B}
\end{align}
This shows that the difference between $E$- and $B$-modes come entirely from the geometric terms, and the latter are in general non-zero even for $B$ modes.

Note that we have separated the groups of indices on which the functions depend fully symmetrically: $A_{\ell_1 \ell_2, \ell_3}(k_1, k_2, k_3)$ and $\widetilde{A}_{\ell_1 \ell_2, \ell_3}(k_1, k_2, k_3)$ are symmetric under exchange of $(\ell_1, k_1)$ with $(\ell_2, k_2)$, $Q_{E\ell_1 \ell_2, \ell_3\ell_4}^{m_1 m_2, m_3 m_4}$ and $\widetilde{Q}_{E\ell_1 \ell_2, \ell_3\ell_4}^{m_1 m_2, m_3 m_4}$ are symmetric under exchange of $(\ell_1, m_1)$ with $(\ell_2, m_2)$, as well as under exchange of $(\ell_3, m_3)$ with $(\ell_4, m_4)$, $Q_{B\ell_1 \ell_2, \ell_3;\ell_4}^{m_1 m_2, m_3; m_4}$ and $\widetilde{Q}_{B\ell_1 \ell_2, \ell_3;\ell_4}^{m_1 m_2, m_3; m_4}$ are symmetric under exchange of $(\ell_1, m_1)$ with $(\ell_2, m_2)$, and anti-symmetric under exchange of $(\ell_3, m_3)$ with $(\ell_4, m_4)$. 

\subsection{Angular Trispectra}

\subsubsection{General form}

We now move on to computing all the connected four-point correlations of temperature and polarization anisotropy, 
\begin{align}
    \langle w_1 x_2 y_3 z_4 \rangle_c &\equiv \langle w_1 x_2 y_3 z_4 \rangle  -\langle w_1x_2\rangle\langle y_3z_4\rangle\nonumber\\
    &- \langle w_1y_3\rangle\langle x_2z_4\rangle - \langle w_1z_4\rangle\langle x_2y_3\rangle, \label{eq:4pt-c}
\end{align}
where $w,x,y,z\in \{\Theta,E,B\}$ and, for short, $w_1\equiv w_{\ell_1 m_1}$, etc... We follow Paper II very closely with a few generalizations.

Recall that each field $x$ consists of $x = x^{(0)} + x^{(1)\rm hom} + x^{(1)\rm inh}$ (with $B^{(0)} = B^{(1)\rm hom} = 0$). Because $x^{(0)}$ and $x^{(1)\rm hom}$ are both linear in the initial Gaussian curvature perturbation, then to lowest order in electron density perturbations, the trispectra are given by
\begin{align}\label{eq:4pt}
    \langle w_1 x_2 y_3 z_4\rangle_c &= \langle w_{1}^{(1)\rm inh} x_2^{(0)} y_3^{(0)} z_4^{(0)} \rangle_c \nonumber\\
    &+  \langle w_1^{(0)} x_{2}^{(1)\rm inh} y_3^{(0)} z_4^{(0)} \rangle_c \nonumber\\
    &+  \langle w_1^{(0)} x_2^{(0)} y_{3}^{(1)\rm inh} z_4^{(0)} \rangle_c \nonumber\\
    &+  \langle w_1^{(0)} x_2^{(0)} y_3^{(0)} z_{4}^{(1)\rm inh} \rangle_c.
\end{align}
We may now compute each term using the general definition of transfer functions in Eq.~\eqref{eq:EB_transf_int}. For instance, the last term results in
\begin{align}
    \langle w_{1}^{(0)} x_2^{(0)} y_3^{(0)} z_{4}^{(1)\rm inh} \rangle_c = \fbh \int D(k k' k'') T_{z4}^{(1)}(\bk, \bk', \bk'') \nonumber\\
    \times \Big{[}\langle \zeta(\bk) \zeta(\bk') \zeta(\bk'') w_1^{(0)} x^{(0)}_2 y_3^{(0)} \rangle \nonumber\\
    - \langle \zeta(\bk) \zeta(\bk') \zeta(\bk'') w_1^{(0)}\rangle \langle x_2^{(0)} y_3^{(0)} \rangle\nonumber\\
    - \langle \zeta(\bk) \zeta(\bk') \zeta(\bk'') x_2^{(0)}\rangle \langle w_1^{(0)} y_3^{(0)} \rangle\nonumber\\
    - \langle \zeta(\bk) \zeta(\bk') \zeta(\bk'') y_3^{(0)}\rangle \langle w_1^{(0)} x_2^{(0)} \rangle\Big{]}.
\end{align}
We can then use Wick's theorem to compute the 6-point and 4-point functions of Gaussian fields appearing in the integrand above, to arrive at,
\begin{align}
\langle w_1^{(0)} x_2^{(0)} y_3^{(0)} z_{4}^{(1)\rm inh} \rangle_c  = \fbh \int D(k k' k'')  \nonumber\\
\times 
\langle \zeta(\bk) w_1^{(0)}\rangle\langle \zeta(\bk') x_2^{(0)}\rangle\langle \zeta(\bk'') y_3^{(0)}\rangle \nonumber\\
\times \left[T_{z4}^{(1)}(\bk, \bk', \bk'') + 5  \textrm{ perms.} \right]
\end{align}
where the 5 permutations involve all other possible permutations of $\bk, \bk', \bk''$. Using Eqs.~\eqref{eq:X0_lm}-\eqref{eq:TX0_lm}, we obtain
\begin{align}
   \langle \zeta(\bk) w_{\ell m}^{(0)}\rangle = 4 \pi (-i)^\ell Y_{\ell m}^*(\hk) \Delta^{(0)}_{w\ell}(k) P_\zeta(k).
\end{align}
Integrating over the wavenumbers' directions, and using the harmonic decomposition of $T_z^{(1)}$ given in Eq.~\eqref{eq:T_multipoles}, we thus arrive at
\begin{align}
    \langle w_{\ell_1 m_1}^{(0)} x_{\ell_2 m_2}^{(0)} y_{\ell_3 m_3}^{(0)} z_{\ell_4 m_4}^{(1)\rm inh} \rangle_c 
    = (4 \pi)^3 \fbh \int D(k_1 k_2 k_3) \nonumber\\
    \times P_\zeta(k_1) P_\zeta(k_2) P_\zeta(k_3) \Delta^{(0)}_{w\ell_1}(k_1) \Delta^{(0)}_{x\ell_2}(k_2) \Delta^{(0)}_{y\ell_3}(k_3) \nonumber\\
   \times \Big{[} T_{z\ell_1 \ell_2 \ell_3 \ell_4}^{m_1 m_2 m_3 m_4}(k_1, k_2, k_3) + 5 \textrm{~perms.} \Big{]},
\end{align}
where the 5 permutations involve all other possible permutations of $k_1, k_2, k_3$ as long as the corresponding indices $\ell_i, m_i, i = 1, 2, 3$ are permuted simultaneously, i.e.~such that the position of the index $\ell_i, m_i$ always corresponds to the position of $k_i$.

In the same vein as Paper~II, we now utilize the factorized forms of $T_{w\ell_1 \ell_2 \ell_3 \ell_4}^{m_1 m_2 m_3 m_4}(k_1, k_2, k_3)$ given in Eqs.~\eqref{eq:Tmult-final}. We define the following functions of time, multipole and field\footnote{Note that all terms which depend on the temperature field (e.g. $\mathcal{J}_{T\ell}$) are defined in Paper~II, though therein they do \textit{not} have the $T$ subscript.}:
\begin{align}
    \lambda^{xy}_{\ell}(\eta) &\equiv \int Dk ~P_\zeta(k) \Delta^{(0)}_{x\ell}(k) \mathcal{J}_{y\ell}(\eta, k), \label{eq:lambda_l}\\
    \mu^x_{\ell}(\eta) &\equiv \int Dk ~P_\zeta(k) \Delta_e(\eta, k) \Delta^{(0)}_{x\ell}(k)  j_\ell'(k \chi) \label{eq:mu_l},\\
    \nu^x_{\ell}(\eta) &\equiv f_\ell \int Dk ~P_\zeta(k) \Delta_e(\eta, k) \Delta^{(0)}_{x\ell}(k)  \frac{j_\ell(k \chi )}{k \chi}. \label{eq:nu_l}
\end{align}

From Eqs.~\eqref{eq:lambda_l}-\eqref{eq:nu_l} we then define the following one-dimensional integrals:
\begin{align}
    \mathcal{A}^{w x y z}_{\ell_1\ell_2\ell_3} \equiv& - 2(4 \pi)^3 \int d\eta ~g(\eta)~ \mu^w_{\ell_1}(\eta) \mu^x_{\ell_2}(\eta) \lambda^{yz}_{\ell_3}(\eta),\label{eq:mathcalA}\\
    \widetilde{\mathcal{A}}^{w x y z}_{\ell_1\ell_2\ell_3} \equiv&  - 2(4 \pi)^3 \int d\eta ~g(\eta)~ \nu^w_{\ell_1}(\eta) \nu^x_{\ell_2}(\eta) \lambda^{yz}_{\ell_3}(\eta),\label{eq:mathcalB}
\end{align}
which are symmetric under exchange of the first two fields $w$ and $x$ simultaneously with their corresponding indices $\ell_1$ and $\ell_2$. We then find, using the symmetry of $T^{(1)}$ in its first two arguments, and the symmetries of the $Q$ and $\widetilde{Q}$ symbols defined in Eqs.~\eqref{eq:Q_sym_E}, \eqref{eq:Qt_sym_E} (with the equivalent ones defined for temperature in Eq.~67,~68 in Paper II): 
\begin{align}
    &\frac1{\fbh}\langle w_{\ell_1 m_1}^{(0)} x_{\ell_2 m_2}^{(0)} y_{\ell_3 m_3}^{(0)} z_{\ell_4 m_4}^{(1)\rm inh} \rangle_c \nonumber\\
    &=\mathcal{A}^{w x y z}_{\ell_1\ell_2\ell_3}{Q}_{ z\ell_1 \ell_2,\ell_3\ell_4}^{m_1 m_2, m_3 m_4}+\widetilde{\mathcal{A}}^{w x y z}_{\ell_1\ell_2\ell_3}\widetilde{Q}_{z \ell_1 \ell_2,\ell_3\ell_4}^{m_1 m_2, m_3 m_4} \nonumber\\
    &\quad+ (w, 1) \leftrightarrow (y, 3)~ + ~(x,2) \leftrightarrow (y, 3).
\end{align}
Using Eq.~\eqref{eq:4pt}, we arrive at the main result of this work, the general temperature and polarization trispectrum sourced by accreting PBHs: 
\begin{align}
    &\langle w_{\ell_1 m_1} x_{\ell_2 m_2} y_{\ell_3 m_3} z_{\ell_4 m_4} \rangle_c =f_{\rm pbh}(\mathcal{T}^{wxyz}_{\rm pbh})_{\ell_1,\ell_2,\ell_3,\ell_4}^{m_1m_2m_3m_4},
\end{align}
with 
\begin{align}\label{eq:verb}
    &(\mathcal{T}^{wxyz}_{\rm pbh})_{\ell_1\ell_2\ell_3\ell_4}^{m_1m_2m_3m_4}\nonumber\\
    &=\Bigg{\{}\Big{[}~~\mathcal{A}^{w x y z}_{\ell_1\ell_2\ell_3}\Qgeo{z}{1}{2}{3}{4}+\mathcal{A}^{w x z y}_{\ell_1\ell_2\ell_4}\Qgeo{y}{1}{2}{4}{3}& \nonumber\\
    &~~~~ +\mathcal{A}^{y z w x}_{\ell_3\ell_4\ell_1}\Qgeo{x}{3}{4}{1}{2}+\mathcal{A}^{y z x w}_{\ell_3\ell_4\ell_2 }\Qgeo{w}{3}{4}{2}{1} \Big{]} \nonumber\\
    & ~~~~ + (\mathcal{A}, Q) \rightarrow (\widetilde{\mathcal{A}}, \widetilde{Q}) \Bigg{\}}\nonumber\\
    &+ (x, 2)\leftrightarrow (y, 3) ~ + ~ (x, 2) \leftrightarrow (z, 4).
\end{align}
This general expression comprises all the unique terms that are not redundant due to the symmetry of all quantities in their first 2 (groups of) indices. 

\subsubsection{Factorization of a unique geometric dependence}

As we will see in the next section, the trispectrum signal-to-noise squared involves summing products of trispectra over all $m$'s. In Paper II, we had explicitly computed the sums over $m$'s of products of all permutation of $Q$ and $\widetilde{Q}$ symbols relevant to the temperature trispectrum, which could be grouped into 5 unique rotationally-invariant quantities. When dealing with temperature and polarization trispectra, this direct approach becomes too cumbersome and more prone to errors. Therefore, instead, we use a more elegant approach here (and cross-check our results from Paper II with this new method), consisting in first projecting all trispectra on the same basis of products of 3-$J$ symbols \citep{hu02a}:
\begin{align}\label{eq:gen_tri}
    &(\mathcal{T}^{wxyz}_{\rm pbh})_{\ell_1\ell_2\ell_3\ell_4}^{m_1m_2m_3m_4} \equiv\sum_{L M} (-1)^M\nonumber\\
    &\quad\quad\times\threej{\ell_1}{\ell_2}{L}{m_1}{m_2}{-M}\threej{L}{\ell_3}{\ell_4}{M}{m_3}{m_4} T^{w x y z}_{\ell_1 \ell_2 \ell_3 \ell_4}(L),
\end{align}
with $w,x,y,z\in \{\Theta,E,B\}$. Indeed, using orthognality relations of 3-$J$ symbols, this expression allows us to easily rewrite the quadruple sum over $m_1, m_2, m_3, m_4$'s of products of tripsectra as a single sum over $L$ of products of reduced trispectra: 
\barr
\sum_{m's}(\mathcal{T}^{abcd}_{\rm pbh})_{\ell_1\ell_2\ell_3\ell_4}^{m_1m_2m_3m_4} (\mathcal{T}^{wxyz}_{\rm pbh})_{\ell_1\ell_2\ell_3\ell_4}^{m_1m_2m_3m_4} \nonumber\\
= \sum_L \frac1{2 L + 1} T^{a b c d}_{\ell_1 \ell_2 \ell_3 \ell_4} (L)T^{w x y z}_{\ell_1 \ell_2 \ell_3 \ell_4} (L). \label{eq:T-squared}
\earr

We explicitly project the $Q$ and $\widetilde{Q}$ symbols on the common 3-$J$ symbol basis in Appendix~\ref{app:Q-sym}, so that our final result for the reduced trispectrum induced by accreting PBHs is
\begin{align}\label{eq:red_tri}
    &{T}^{wxyz}_{\ell_1 \ell_2 \ell_3 \ell_4}(L)=\nonumber\\
    &\left\{\mathcal{A}^{w x y z}_{\ell_1\ell_2\ell_3}\alpha^{\ell_1 \ell_2 \ell_3 \ell_4}_{zL}\! +\!\mathcal{A}^{w x z y}_{\ell_1\ell_2\ell_4}\alpha^{\ell_1 \ell_2 \ell_4 \ell_3}_{yL}(-1)^{L + \ell_{34}} \right.\nonumber\\
    &+\mathcal{A}^{y z w x}_{\ell_3\ell_4\ell_1}\alpha^{\ell_3 \ell_4 \ell_1 \ell_2}_{xL} +\mathcal{A}^{y z x w}_{\ell_3\ell_4\ell_2}\alpha^{\ell_3 \ell_4 \ell_2 \ell_1}_{wL}(-1)^{L + \ell_{12}}\nonumber\\
    &+\mathcal{A}^{w y x z}_{\ell_1\ell_3\ell_2}\beta^{\ell_1 \ell_3 \ell_2 \ell_4}_{zL}+\mathcal{A}^{x z w y}_{\ell_2\ell_4\ell_1}\beta^{\ell_2 \ell_4 \ell_1 \ell_3}_{yL}(-1)^{\ell_{1234}}\nonumber\\
    &+\mathcal{A}^{w z x y}_{\ell_1\ell_4\ell_2}\beta^{\ell_1 \ell_4 \ell_2 \ell_3}_{yL}(-1)^{L + \ell_{34}}+\mathcal{A}^{y x z w}_{\ell_3\ell_2\ell_4}\beta^{\ell_3 \ell_2 \ell_4 \ell_1}_{wL}(-1)^{L + \ell_{12}}\nonumber\\
    &+\mathcal{A}^{w y z x}_{\ell_1\ell_3\ell_4}\gamma^{\ell_1 \ell_3 \ell_4 \ell_2}_{xL}+\mathcal{A}^{x z y w}_{\ell_2\ell_4\ell_3}\gamma^{\ell_2 \ell_4 \ell_3 \ell_1}_{wL}(-1)^{\ell_{1234}}\nonumber\\
    &\left.+\mathcal{A}^{w z y x}_{\ell_1\ell_4\ell_3}\gamma^{\ell_1 \ell_4 \ell_3 \ell_2}_{xL}(-1)^{L + \ell_{34}}+\mathcal{A}^{y x w z}_{\ell_3\ell_2\ell_1}\gamma^{\ell_3 \ell_2 \ell_1 \ell_4}_{zL}(-1)^{L + \ell_{12}}\!\right\}\nonumber\\
    &+(\mathcal{A}, \alpha,\beta,\gamma)\rightarrow(\widetilde{\mathcal{A}}, \widetilde{\alpha},\widetilde{\beta},\widetilde{\gamma}),
\end{align}
where $\ell_{12} \equiv \ell_1+\ell_2$, $\ell_{34} \equiv \ell_3+\ell_4$, $\ell_{1234} \equiv \ell_1 + \ell_2 + \ell_3 + \ell_4$, and $\alpha_L,\beta_L,\gamma_L$ and their tilde counterparts are defined in Table~\ref{Tab:coeffs}.

\begin{table*}[ht]
\begingroup
\renewcommand{\arraystretch}{2}
\begin{tabular}{|c||c|c|c|}
\hline
     $x$ & $T$ & $E$ & $B$  \\
     \hline\hline
     $\alpha_{x L}^{\ell_1 \ell_2 \ell_3 \ell_4}$ & $g_{\ell_1\ell_2 L} ~g_{\ell_3 \ell_4 L}$ & $g_{\ell_1 \ell_2 L}~ g_{\ell_3 \ell_4 L}^{2, -2, 0} \mathfrak{E}(L + \ell_{34})$ & $-i ~g_{\ell_1 \ell_2 L}~ g_{\ell_3 \ell_4 L}^{2, -2, 0} \mathfrak{O}(L + \ell_{34})$  \\
     
     $\widetilde{\alpha}_{x L}^{\ell_1 \ell_2 \ell_3 \ell_4}$ & $-g_{\ell_1 \ell_2 L}^{1, -1, 0} g_{\ell_3 \ell_4 L} ~\mathfrak{E}(L+\ell_{12})$ & $-g^{1,-1,0}_{\ell_1\ell_2 L}g^{2,-2,0}_{\ell_3\ell_4 L} \mathfrak{E}(L+\ell_{12}) \mathfrak{E}(L + \ell_{34})$ & $i~ g^{1,-1,0}_{\ell_1\ell_2 L}g^{2,-2,0}_{\ell_3\ell_4 L} \mathfrak{E}(L+\ell_{12}) \mathfrak{O}(L + \ell_{34})$ \\
     
     $\beta_{x L}^{\ell_1 \ell_2 \ell_3 \ell_4}$ & $g_{\ell_1\ell_3 L} ~g_{\ell_2 \ell_4 L}$ & $g_{\ell_1 \ell_3 L}^{0, 2, -2} g_{\ell_2 \ell_4 L}^{0, -2, 2} \mathfrak{E}(\ell_{1234})$ & $-i~g_{\ell_1 \ell_3 L}^{0, 2, -2} g_{\ell_2 \ell_4 L}^{0, -2, 2} \mathfrak{O}(\ell_{1234})$ \\
     
     $\widetilde{\beta}_{x L}^{\ell_1 \ell_2 \ell_3 \ell_4}$ & $g^{-1,0,1}_{\ell_1\ell_3 L}g^{1,0,-1}_{\ell_2\ell_4 L}\mathfrak{E}(\ell_{1234})$& \!$\frac12 (g_{\ell_1 \ell_3 L}^{-1, 2, -1} g_{\ell_2 \ell_4 L}^{1, -2, 1} + g_{\ell_1 \ell_3 L}^{1 , 2, -3} g_{\ell_2 \ell_4 L}^{-1, -2, 3}) \mathfrak{E}(\ell_{1234})$\! &\!\!\! $-\frac{i}2  (g_{\ell_1 \ell_3 L}^{-1, 2, -1} g_{\ell_2 \ell_4 L}^{1, -2, 1} + g_{\ell_1 \ell_3 L}^{1 , 2, -3} g_{\ell_2 \ell_4 L}^{-1, -2, 3}) \mathfrak{O}(\ell_{1234})$\! \\[2pt]
     \hline
\end{tabular} 
\caption{Projection coefficients of the geometric $Q$ and $\widetilde{Q}$ symbols, defined in Eq.~\eqref{eq:alpha def}-\eqref{eq:beta def}. We recall that $g_{\ell_1 \ell_2 \ell_3}^{s_1 s_2 s_3}$ is defined in Eq.~\eqref{eq:g_sym}, and define for compactness $g_{\ell_1 \ell_2 \ell_3} \equiv g_{\ell_1 \ell_2 \ell_3}^{0\, 0\, 0}$, as well as $\ell_{ij...} \equiv \ell_i + \ell_j + ..$. The symbol $\mathfrak{E}(\ell)$ is 1 if $\ell$ is even, and 0 otherwise, and $\mathfrak{O}(\ell) = 1 - \mathfrak{E}(\ell)$ is 1 if $\ell$ is odd and 0 otherwise. Note that $g_{\ell_1 \ell_2 \ell_3} \propto \mathfrak{E}(\ell_{123})$. The $T$ and $E$ coefficients always vanish when the $B$ coefficient is non-zero and vice-versa. Moreover, due to symmetries, $\gamma^{\ell_1 \ell_2 \ell_3 \ell_4}_{\Theta/EL}=\beta^{\ell_1 \ell_2 \ell_4 \ell_3}_{\Theta/EL}$ and $\gamma^{\ell_1 \ell_2 \ell_3 \ell_4}_{BL}=-\beta^{\ell_1 \ell_2 \ell_4 \ell_3}_{BL}$.}\label{Tab:coeffs}
\endgroup
\end{table*}

\section{Constraints and sensitivity forecasts}\label{sec:forecast}
In this section we compute sensitivity forecasts on the fraction of dark matter made of PBHs, $f_{\rm pbh}$. We use $T\leftrightarrow \Theta$ interchangeably. 

\subsection{General considerations and equations}

We utilize every nonzero combination of fields $T$, $E$, and $B$ in the first-order trispectrum (linear in $f_{\rm pbh}$). For the scope of this paper, we neglect other nonlinear effects like lensing, hence assume that $B$ modes vanish in the unperturbed $\Lambda$CDM model. It was shown in Section~\ref{sec:trispec} that, when introducing an inhomogeneous free-electron fraction, $B$ modes were induced. As a consequence, there are non-vanishing trispectra involving one $B$ mode at linear order in $f_{\rm pbh}$. Note that these trispectra do not vanish (contrary to the parity-odd cross-power spectra $\av{T B}$ and $\av{EB}$); instead, they are antisymmetric under parity transformations in real space, similarly to $B$-mode bispectra \cite{Meerburg_16}. Thus, at linear order in $f_{\rm pbh}$, there are nine unique tripectra: the five parity-even tripectra
\beq
\av{T T T T}_c, \av{TTTE}_c, \av{TTEE}_c, \av{TEEE}_c, \av{EEEE}_c,
\eeq
and the four parity-odd trispectra
\beq
\av{TTTB}_c, \av{TTEB}_c, \av{TEEB}_c, \av{EEEB}_c.
\eeq

Similar to the temperature-only case of the trispectrum discussed in Paper~II, one can construct an optimal quartic estimator $\widehat{f}_{\rm pbh}$ incorporating all combinations of non-zero first-order trispectra \citep{fergusson14a, okamoto02a}. We do not use the explicit expression in this paper, but an example can be found in Eq.~(A4) in Ref.~\citep{fergusson14a}. In practice we compute the inverse variance of the general estimator given by Eq.~(27) of Ref.~\citep{okamoto02a}. We approximate the covariance matrix as diagonal in $\ell$, resulting in the variance of our estimator,
\begin{align}
    \frac{1}{\sigma^2_{f_{\rm pbh}}}=&\frac{f_{\rm sky}}{4!}\sum_{abcd}\sum_{wxyz}\sum_{\ell's}\nonumber\\
    &\times(C^{'-1})^{aw}_{\ell_1}(C^{'-1})^{bx}_{\ell_2}(C^{'-1})^{cy}_{\ell_3}(C^{'-1})^{dz}_{\ell_4}\nonumber\\
    &\times \sum_{m's}(\mathcal{T}^{abcd}_{\rm pbh})_{\ell_1\ell_2\ell_3\ell_4}^{m_1m_2m_3m_4}(\mathcal{T}^{*wxyz}_{\rm pbh})_{\ell_1\ell_2\ell_3\ell_4}^{m_1m_2m_3m_4},\nonumber\\
    =&\frac{f_{\rm sky}}{4!}\sum_{abcd}\sum_{wxyz}\sum_{\ell's}\nonumber\\
    &\times(C^{'-1})^{aw}_{\ell_1}(C^{'-1})^{bx}_{\ell_2}(C^{'-1})^{cy}_{\ell_3}(C^{'-1})^{dz}_{\ell_4}\nonumber\\
    &\times \sum_L \frac1{2 L + 1} T^{a b c d}_{\ell_1 \ell_2 \ell_3 \ell_4} (L)T^{w x y z}_{\ell_1 \ell_2 \ell_3 \ell_4} (L),\label{eq:invvar}
\end{align}
where in the second line we used Eq.~\eqref{eq:T-squared} to convert the quadruple sum over $m$'s into a single sum over $L$. In this equation, $f_{\rm sky}$ is the fraction of the sky covered by the experiment, $a,b,c,d,w,x,y,z\in\{T,E,B\}$ are summed over every trispectrum field configuration, and $C^{'aw}_{\ell}\equiv C^{aw}_{\ell}+N^{aw}_{\ell}$ is the total variance between two fields $a$ and $w$, including both the cosmological signal $C^{aw}_{\ell}$ and instrumental noise $N^{aw}_{\ell}$. The inverse, $(C^{'-1})^{aw}_{\ell}$, is the $aw$ element of the inverse of the covariance matrix,
\begin{align}
C_\ell^{'-1}\equiv
    \begin{pmatrix}
        C^{'TT}_{\ell} & C^{'TE}_{\ell} & 0 \\
        C^{'ET}_{\ell} & C^{'EE}_{\ell} & 0 \\
        0 & 0 & C^{'BB}_{\ell}
    \end{pmatrix}^{-1}.
\end{align}
In an ideal experiment, the instrumental noises of polarization and temperature are uncorrelated, hence we take $C^{'TE}_\ell=C_\ell^{TE}$. We shall moreover neglect $C_{\ell}^{\rm BB}$ and assume that $C^{'BB}_{\ell}\approx N^B_{\ell} = N_\ell^E$. This is an excellent approximation for Planck, since the lensing $B$-mode signal is subdominant to Planck's noise at all multipoles. For a CMB-S4 experiment, we are assuming that the lensed $B$ modes can be efficiently subtracted by delensing.

Note that if we considered only the temperature trispectrum, the form of Eq.~\eqref{eq:invvar} is different than (although equivalent to) Eq.~(113) in Paper~II, thus resulting in using different computational methods in this paper discussed in Appendix~\ref{app:comp}. We checked explicitly that our current approach identically recovers the results we obtained in Paper~II for the temperature trispectrum. Moreover in Appendix~\ref{app:comp}, we explored a third independent computational method which reproduces identical results when also incorporating polarization data, giving us confidence in all three methods and in our numerical implementation.

\subsection{Application to Planck and CMB-S4}\label{subsec:noise}

We now apply the above results to an ideal Planck experiment \citep{planck20a,planck20b,planck20c} and to a CMB Stage-4-like experiment \cite{CMB-S4:2016}.

For Planck, the relevant fraction of sky coverage is $f_{\rm sky}=0.78$ \citep{planck20c}, and the instrumental noise $N^{x}_\ell$ for temperature ($x=T$) or polarization ($x=E,B$) is obtained from combining the noises of the 100, 143 and 217 GHz frequency channels, 
\begin{align}
    N^x_\ell=\left[\sum_c N^{x-1}_{\ell,c}\right]^{-1},
\end{align}
where, for each channel $c$, the noise is modelled as a Gaussian with variance per pixel $\sigma^2_c$ and beam size $\theta_{{\rm FWHM},c}$:
\begin{align}
    N^x_{\ell,c}=\left(\frac{\sigma^x_c~\theta_{{\rm FWHM},c}}{T_0}\right)^2\exp\left[\frac{\ell(\ell+1)\theta^2_{{\rm FWHM},c}}{8\ln 2}\right],
\end{align}
where $T_0=2.73$ K is the CMB monopole. The respective parameters for each channel are in the following table\footnote{\url{https://wiki.cosmos.esa.int/planckpla/index.php/Main_Page}. N.b. for polarization we use the goal sensitivity of Planck from Ref.~\citep{planckbluebook}.}:
\begin{center}
\begin{tabular}{ c c c c }
 $\nu_c$(GHz) & $\theta_{\rm FWHM, c}$ & $\sigma^T_c (\mu$K) & $\sigma^{E,B}_c (\mu$K) \\ 
 \hline
  100 & 9.66$'$ & 10.77 & 10.90 \\ 
  143 & 7.27$'$ & 6.40  & 11.45 \\ 
  217 & 5.01$'$ & 12.48 & 26.71 \\ 
\end{tabular}
\end{center}

For a CMB Stage-4-like experiment \cite{CMB-S4:2016}, we assume $f_{\rm sky} = 0.4$ and a single effective frequency with beam size $\theta_{\rm FWHM} = 1'$ and noises $\sigma^T = 1 ~\mu$K, $\sigma^{E, B} = \sqrt{2} ~\mu$K.

\subsection{Results and discussion}
We are now in the position to forecast the sensitivity of Planck and of a CMB-S4 experiment to accreting PBHs through their non-Gaussian signatures. We follow the same procedure as in Paper~II, and compare our new results to the Planck power-spectra limits on $f_{\rm pbh}$, using Planck 2018 data \citep{planck20a}. To reiterate, to extract power-spectra limits we Taylor-expand near the Planck best-fit cosmology using the foreground-marginalized Plik-lite log-likelihood for $C_\ell$'s at $\ell \geq 30$. We account approximately for the low-$\ell$ data by imposing a Gaussian prior on the optical depth to reionization $\tau_{\rm re}$. When presenting power-spectrum forecasts for CMB-S4, we use a Fisher analysis, using a Gaussian Likelihood for $C_\ell's$ with $\ell \geq 30$ and the same Gaussian prior for $\tau_{\rm re}$ (we checked that using a Gaussian likelihood for the $C_\ell$'s with all $\ell \geq 2$ makes no significant difference).

\begin{figure*}[ht]
\includegraphics[width=2\columnwidth]{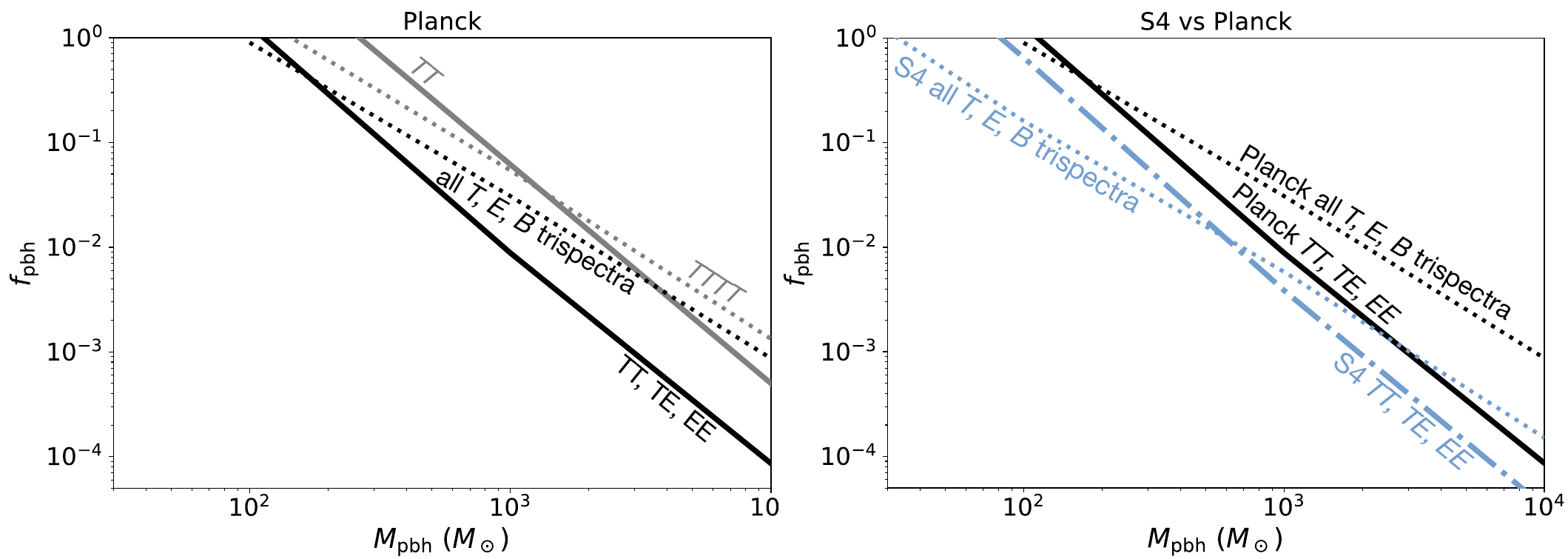}
\caption{\label{fig:const} \emph{Left}: Planck 2018 CMB 2-point-function constraints (solid lines) and 4-point-function sensitivity forecasts (dashed lines) to the fraction of dark matter in PBHs, as a function of PBH mass. Gray lines only include temperature and black lines include temperature and polarization. Planck's forecasted sensitivity when using all temperature and polarization trispectra is better than current-day power-spectrum constraints only for $M_{\rm pbh} \lesssim 150~ M_{\odot}$. \emph{Right}: Comparison between Planck (black lines) and CMB-S4 (blue lines). A CMB-S4 experiment would be more sensitive than Planck by a factor $\sim 3$ for a power spectrum analysis, and by a factor $\sim 6$ for a trispectrum analysis. While in the case of Planck data, $B$-mode trispectra only marginally contribute to the forecasted sensitivity, for a CMB-S4 experiment, their inclusion more than doubles it. For a CMB-S4 experiment, a 4-point function analysis would significantly improve sensitivity over a 2-point function analysis.}
\end{figure*}

The left panel of Fig.~\ref{fig:const} shows Planck's forecasted 1-$\sigma$ sensitivity to PBHs through the 4-point functions they induce. We compare these results with current Planck power-spectra upper limits on $f_{\rm pbh}$. We find that Planck's forecasted sensitivity from all trispectra is competitive with temperature and polarization power-spectra limits only for $M_{\rm pbh} \lesssim 150~ M_{\odot}$. The forecasted 4-point function sensitivity is thus not nearly as good as what we had anticipated from the simple order-of-magnitude arguments given in the introduction. Note that throughout we assumed the most conservative ``collisional ionization" limit of AK17, but checked that our qualitative conclusions remain unchanged for the ``photoionization" limit (see AK17 for details). Below we give some post-factum arguments that partially make sense of our numerical result.

As shown in the left panel of Fig.~\ref{fig:const}, adding polarization trispectra (including both $E$ and $B$ modes) only increases the forecasted sensitivity by a factor $\sim 2$ over the $TTTT$-only sensitivity, be it in the collisional ionization or photoionization limits. This is to be contrasted with the order-of-magnitude improvement in constraints going from the $TT$-only power spectrum to incorporating all $TT,TE,EE$ power spectra data. This can be understood as follows: when considering 2-point statistics, incorporating polarization data breaks the degeneracy between perturbed recombination and the optical depth to reionization, leading to a significant boost in sensitivity as compared to a temperature-only analysis. There is no such degeneracy to be broken in the trispectra analysis, especially as we neglect secondary effects that could also contribute to non-Gaussianities. To drive this point further, we hold all cosmological parameters constant and repeat a power-spectra sensitivity forecast; we find that incorporating polarization would only improve temperature-only constraints by $\sim 10\%$ in this unrealistic, degeneracy-free 2-point-function analysis.

Surprisingly, we found that the factor-2 improvement in Planck's sensitivity from polarization trispectra is almost entirely due to non-Gaussianity in the temperature rather than in the polarization, i.e.~from terms such as $\av{T^{(1)} T^{(0)} T^{(0)} E^{(0)}}$ and $\av{T^{(1)} T^{(0)} E^{(0)} E^{(0)}}$ rather than $\av{T^{(0)} T^{(0)} T^{(0)} E^{(1)}}$ and $\av{T^{(0)} T^{(0)} E^{(0)} E^{(1)}}$ or their $B$-mode counterparts. Indeed, we found that neglecting the contributions of $E^{(1)}$ and $B^{(1)}$ to the trispectra only changed the forecasted Planck sensitivity by $\sim 10\%$. We have not been able to find an intuitive argument explaining this unexpected result. It is possible that some geometric considerations may lead to weaker 4-point functions involving perturbed polarization fields, qualitatively similar to the vanishing of 2-point correlation functions involving them. However, these 4-point functions clearly do not vanish identically, and we have not been able to identify any fundamental reason for their smallness. We have thoroughly examined various intermediate quantities, extensively tested our code, and even implemented a second code with a completely different algorithm (see Appendix~\ref{app:comp}), and found identical outcomes. We are therefore confident in our numerical results, despite their being somewhat counter-intuitive. 

The right panel of Fig.~\ref{fig:const} shows the forecasted sensitivity of a CMB S4-like experiment to accreting PBHs, from both 2-point and 4-point functions. We see that a 2-point-function analysis with a CMB-S4 experiment would be more sensitive than Planck's current limits by a factor $\sim 3$. Excitingly, a 4-point-function analysis with CMB-S4 would be more sensitive than with Planck by a factor of $\sim 6$. It would moreover surpass the sensitivity of a 2-point function search for a broader range of PBH masses ($M_{\rm pbh} \lesssim 600~ M_{\odot}$), and extend by a factor $\sim 3$ the range of PBH masses for which sub-DM abundance ($f_{\rm pbh} \leq 1$) can be probed. Interestingly, the addition of $B$-mode trispectra contribute at least half of a CMB-S4 experiment's sensitivity to 4-point functions sourced by PBHs.

\section{Conclusions}\label{sec:conc}

This is the final installment of a series of three papers calculating the non-Gaussian signatures of inhomogeneously-accreting PBHs in CMB anisotropies. Paper I \citep{jensen21a} rigorously set the groundwork to inspect how inhomogeneous energy injection alters the cosmological ionization history. Paper II \citep{jensen23a} then used this machinery to compute the perturbations to temperature anisotropy and its statistics, ultimately computing the CMB temperature trispectrum and its detectability by Planck. In the present analysis, we conclude the series by computing all 2-point and 4-point statistics involving CMB temperature, $E$- and $B$-mode polarization, and forecasting the sensitivity of Planck and a CMB Stage-4 experiment to temperature-polarization trispectra induced by PBHs.  

We generalized the formalism of Paper II to polarization, promoting the photon distribution function to a tensor-valued one in the Boltzmann-Einstein equations. When perturbing these equations with an inhomogeneous free-electron fraction, both $E$ and $B$ polarization modes are produced, even in the absence of primordial tensor modes. As in previous works, our perturbative calculation only accounts for the direct effect of free-electron perturbations, and neglects the ``feedback" effect, which would require solving a full perturbed Boltzmann hierarchy. Going beyond previous works, however, we explicitly cross-checked this approximation in the case of a homogeneous free-electron perturbation, for which standard Boltzmann codes can output the exact result. We found that our approximation is in excellent agreement with the output from \texttt{CLASS}, giving us confidence in our formalism and approximations.

In the context of accreting PBHs, we computed the perturbations to CMB power spectra due to the inhomogeneous part of the free-electron fraction, but found that any correlation involving the perturbed $E$ mode and an unperturbed field vanishes in the limit of a spatially on-the-spot energy deposition. This implies that current CMB power-spectra limits are unaffected by inhomogeneities in the free-electron perturbation. For the first time, we also computed the $BB$ power spectrum produced by accreting PBHs, which is quadratic in the PBH abundance. We found that this signal is three orders of magnitude below the standard lensing predictions when saturating current CMB-anisotropy limits to PBHs, implying that the $B$-mode power spectrum is not a competitive probe of PBHs. 

The main result of this paper are the trispectra involving all combinations of $T$, $E$, and $B$ induced by accreting PBHs. As opposed to the temperature-only analysis considering the $TTTT$ trispectrum alone done in Paper II, we computed and included nine unique field configurations of trispectra at linear order in the abundance of accreting PBH. 

We forecasted the sensitivity of Planck to these trispectra for the first time. Although our numerical results show weaker sensitivities than what we expected a priori, we still find that for $M_{\rm pbh}\lesssim 150~M_{\odot}$ and for the most conservative accretion model of AK17, Planck would be sensitive to smaller PBH abundances through their 4-point function signal than through their effect on the 2-point functions. We argued that the marginal increase in sensitivity when incorporating polarization trispectra may be partly due to the fact that, in contrast with the 2-point-function case, inclusion of polarization data does not break degeneracies in a 4-point-function analysis. Due to the high dimensionality of trispectra, it is difficult to make simple quantitative arguments fully justifying our numerical outcomes. Nevertheless, we built confidence in our results not only with thorough checks, but also by developing two separate codes using distinct numerical methods, and verifying their exact agreement. Given that we forecast at best a marginal improvement of sensitivity to accreting PBHs when using 4-point functions instead of 2-point functions, and that our analysis did not account for any possible contaminants to the signal (such as CMB lensing), we conclude that it is not worth searching for trispectra induced by accreting PBHs in Planck data.

We also forecasted the sensitivity of a CMB-Stage-4-like experiment to the non-Gaussian signal sourced by accreting PBHs. In contrast with Planck, we found that CMB trispectra, and in particular $B$-mode trispectra, would provide a notable enhancement of sensitivity to accreting PBHs relative to a power-spectrum analysis, extending by a factor of $\sim 3$ the range of masses for which PBH abundances below the total dark matter abundance can be probed. This promising result warrants understanding the contamination of the signal by lensing and other non-Gaussian foregrounds in more detail.

In closing, we provided a very general formalism for the response of CMB polarization to arbitrary inhomogeneous free-electron fraction perturbations, which should be useful beyond the context of PBHs. At a higher level, this is one of the very few works considering CMB signatures of non-standard physics (besides primordial non-Gaussianity) beyond the 2-point functions, and it is our hope that it will inspire similar endeavors: there is more to the primary CMB than its power spectra!

\section*{Acknowledgements}
YAH is a CIFAR-Azrieli Global scholar, and acknowledges funding from CIFAR and from NASA grant 80NSSC20K0532. We thank Nanoom Lee for cross-checking our sensitivity estimate for the CMB-Stage-4 2-point function analysis.

\newpage

\begin{appendix}
\section{Spin-2 plane-wave expansion} \label{app:spin}

The goal of this appendix is to derive the spin-2 equivalent of the scalar plane-wave expansion.

We start by recalling that spin-2 spherical harmonics can be written as projections of double gradients of scalar spherical harmonics on the helicity basis:
\begin{align}
    Y^{\pm 2}_{\ell m}(\hn)= 2 \sqrt{\frac{(\ell-2)!}{(\ell+2)!}}~\epsilon^i_{\pm} \epsilon^j_{\pm}~\nabla_{\hn}^i\nabla_{\hn}^jY_{\ell m}(\hn), \label{eq:Ypm2-def}
\end{align}
where $\bs{\nabla}_{\hn}$ is the covariant derivative on the sphere, and for any vector tangent to the sphere we denote $V^i = V_i$ its components on an orthonormal basis. To compute the double dot product above, it will be convenient to use a global, 3-dimensional orthonormal basis. Our first task is thus to express $\nabla_{\hn}^i$ on such a non-tangent basis.

For a scalar field $f(\hn)$ on the sphere, one can show that
\beq
\nabla_{\hn}^i f(\hn) = \delta^{ij}_{\bot}~ \partial_{n^j}f(\hn), 
\eeq
where $n^j$ are the components of $\hn$ on the global 3-d orthonormal basis, and $\delta^{ij}_{\bot} \equiv \delta^{ij} - n^i n^j$. The result above is rather intuitive, and can be derived by considering, for instance, an extension of $f$ to 3d space, computing its gradient, and evaluating the result on the sphere.

We next seek the covariant derivative of a vector field $\bs{V}(\hn)$ tangent to the sphere in the form
\beq
\nabla_{\hn}^i V^j = \delta^{ik}_{\bot}\partial_{n^k} V^j + \Gamma^{ij}_k V^k,
\eeq
with $\Gamma^{ij}_k n^k = 0$ without loss of generality. This implies that, for any tangent tensor field $\bs{T}(\hn)$, we have
\beq
\nabla_{\hn}^i T^{jk} = \delta^{il}_{\bot}\partial_{n^l} T^{jk} + \Gamma^{ij}_l T^{lk} + \Gamma^{ik}_l T^{jl}.
\eeq
We require this covariant derivative to be torsion-free, i.e.~$\nabla_{\hn}^{[i} \nabla_{\hn}^{j]} f = 0$. After some algebra, we find that this implies that the antisymmetric part of $\Gamma^{ij}_k$ (in its upper two indices) must satisfy 
\beq
\Gamma^{[ij]}_k = \delta^{k[i}_{\bot} n^{j]}. \label{eq:Gamma-asym}
\eeq
Next we require metric compatibility, i.e.~that the covariant derivative applied to the metric tensor $\bs{g}$ vanishes. In the global 3d coordinate system, the components of the metric tensor are just $g^{ij} = \delta^{ij}_{\bot}$. Therefore, we require
\barr
0 &=& \nabla^i_{\hn} g^{jk} = \delta^{il}_{\bot}\partial_{n^l} \delta^{jk}_{\bot} + \Gamma^{ij}_l \delta^{lk}_{\bot} + \Gamma^{ik}_l \delta^{il}_{\bot}\nonumber\\
&=& - \delta^{ij}_{\bot} n^{k} - \delta^{ik}_{\bot} n^{j}   + \Gamma^{ij}_k + \Gamma^{ik}_j.
\earr
We now compute the following combination:
\barr
0 &=& \nabla^{(i}_{\hn} g^{j)k} - \frac12 \nabla^k_{\hn} g^{ij} \nonumber\\
&=& - \delta^{ij}_{\bot} n^{k} +  \Gamma^{(ij)}_k + \Gamma^{[ik]}_j + \Gamma^{[jk]}_i.
\earr
Inserting Eq.~\eqref{eq:Gamma-asym}, we arrive at
\beq
\Gamma_{(ij)}^k = \delta^{ij}_{\bot} n^{k} - \delta^{j[i}_{\bot} n^{k]} - \delta^{i[j}_{\bot} n^{k]} = \delta^{k(i}_\bot n^{j)}.
\eeq
Hence, we conclude that 
\beq
\Gamma^{ij}_k = \delta^{ki}_\bot n^j.
\eeq
Armed with this result, we may now evaluate
\begin{align}
&\nabla_{\hn}^i \nabla^j_{\hn} e^{-i\chi \bk \cdot \hn} =\nonumber\\
&\quad-i \chi k \left( -i \chi k  \hk_\bot^i \hk_\bot^j - \mu \delta^{ij}_\bot+ 2 n^{(i}k^{j)}_\bot\right) e^{-i\chi \bk \cdot \hn},
\end{align}
where $\mu \equiv \hk \cdot \hn$ and $\hk_\bot \equiv \hk - \mu \hn$ is the component of $\hk$ perpendicular to $\hn$. Given that $\bm{\epsilon}_\pm \cdot \hn = 0 = \bm{\epsilon}_{\pm} \cdot \bm{\epsilon}_{\pm}$, this implies
\begin{align}
\epsilon^i_{\pm}\epsilon^j_{\pm} \hk_i \hk_j e^{-i\chi \bk \cdot \hn} = - \frac1{(\chi k)^2} \epsilon^i_{\pm}\epsilon^j_{\pm} \nabla^i_{\hn} \nabla^j_{\hn} e^{-i\chi \bk \cdot \hn}.
\end{align}
Thus, after using the scalar plane-wave expansion Eq.~\eqref{eq:plane-wave} and combining with the definition of spin-2 spherical harmonics Eq.~\eqref{eq:Ypm2-def}, we arrive at the spin-2 plane-wave expansion given in Eq.~\eqref{eq:plane-wave-spin2}.

\section{Q-symbol projections} \label{app:Q-sym}

This appendix is dedicated to re-expressing the geometric $Q$-symbols in a form amenable to efficient computation of the trispectrum SNR$^2$. 

We will make abundant use of the following product rule for spin-weighted spherical harmonics \citep{Qtheory}:
\begin{align}\label{eq:product_rule}
    Y_{\ell_1 m_1}^{s_1}(\hat{n}) Y_{\ell_2 m_2}^{s_2}(\hat{n}) =\sum_{s_3,\ell_3m_3}g^{-s_1(-s_2)(-s_3)}_{\ell_1\ell_2\ell_3}\nonumber\\
    \times\threej{\ell_1}{\ell_2}{\ell_3}{m_1}{m_2}{m_3} Y^{s_3*}_{\ell_3 m_3}(\hat{n}),
\end{align}
where the $g$-symbols are defined by
\begin{align}\label{eq:g_sym}
    g_{\ell_1 \ell_2 \ell_3}^{s_1 s_2 s_3} &\equiv \sqrt{\frac{(2 \ell_1 +1)(2 \ell_2 +1)(2 \ell_3 +1)}{4 \pi}}  \threej{\ell_1}{\ell_2}{\ell_3}{s_1}{s_2}{s_3}.
\end{align}
This allows us to rewrite general integrals of four spin-weighted spherical harmonics as follows (using $Y_{\ell m}^{s*} = (-1)^{\ell + s} Y_{\ell, -m}^{-s}$):
\barr
&&\int\! d^2 \hat{n}~ Y^{-s_1*}_{\ell_1 m_1}(\hat{n}) Y^{-s_2 *}_{\ell_2 m_2}(\hat{n}) Y^{-s_3*}_{\ell_3 m_3}(\hat{n})Y^{-s_4*}_{\ell_4 m_4}(\hat{n}) \nonumber\\
    &=&\sum_{L,M, S} (-1)^M \threej{\ell_1}{\ell_2}{L}{m_1}{m_2}{-M} \threej{L}{\ell_3}{\ell_4}{M}{m_3}{m_4} \nonumber\\
    &&\times (-1)^S g_{\ell_1 \ell_2 L}^{s_1 s_2 (-S)} g_{ \ell_3 \ell_4L }^{s_3 s_4 S},\label{eq:alph_gen}
\earr
where $\ell_{1234} \equiv \ell_1 + \ell_2 + \ell_3 + \ell_4$. Note that the sums over $M$ and $S$ really only have one element, $M = m_1 + m_2 = -(m_3 + m_4)$ and $S = s_1 + s_2 = -(s_3 + s_4)$.

We now apply this procedure to rewrite the geometric $Q$-symbols given in the main text for polarization (c.f. Eq.~\eqref{eq:Q_sym_E}~--~\eqref{eq:Qt_sym_B}) and in Paper II for temperature. We will write each $Q$ symbol in three different forms, corresponding to 3 different possible index pairings:
\begin{align}
&Q_{z\ell_1 \ell_2 \ell_3 \ell_4}^{m_1 m_2 m_3 m_4} \nonumber\\
&= \sum_{L M} (-1)^m \threej{\ell_1}{\ell_2}{L}{m_1}{m_2}{-M}\threej{L}{\ell_3}{\ell_4}{M}{m_3}{m_4} \alpha^{\ell_1 \ell_2 \ell_3 \ell_4}_{zL}, \label{eq:alpha def} \\
&= \sum_{L M} (-1)^m \threej{\ell_1}{\ell_3}{L}{m_1}{m_3}{-M}\threej{L}{\ell_2}{\ell_4}{M}{m_2}{m_4} \beta^{\ell_1 \ell_2 \ell_3 \ell_4}_{zL}, \label{eq:beta def}\\
&= \sum_{L M} (-1)^m \threej{\ell_1}{\ell_4}{L}{m_1}{m_4}{-M}\threej{L}{\ell_2}{\ell_3}{M}{m_2}{m_3} \gamma^{\ell_1 \ell_2 \ell_3 \ell_4}_{zL}, \label{eq:gamma def}
\end{align}
and identically for $\widetilde{Q}$-symbols defined with $\widetilde{\alpha},\widetilde{\beta},\widetilde{\gamma}$. Given that all $Q$ symbols are symmetric under exchange of the first 2 $(\ell, m)$ pairs of indices, we must have
\begin{align}
\alpha^{\ell_2 \ell_1 \ell_3 \ell_4}_L =& (-1)^{L + \ell_{12}} ~\alpha^{\ell_1 \ell_2 \ell_3 \ell_4}_L, \ \ \ \ell_{12} \equiv \ell_1 + \ell_2,\\
\gamma_L^{\ell_1 \ell_2 \ell_3 \ell_4} =& \beta_L^{\ell_2 \ell_1 \ell_3 \ell_4}.
\end{align}
Hence we see that $\alpha$ and $\beta$ coefficients are enough to fully describe the $Q$ symbols. 

Given that the $Q$ and $\tilde{Q}$ symbols associated with $T$ and $E$ are symmetric under exchange of their last 2 pairs of indices, and those associated with $B$ are antisymmetric under the same operation, we must have
\barr
\alpha^{\ell_1 \ell_2 \ell_4 \ell_3}_{T/EL} &=& (-1)^{L + \ell_{34}} ~\alpha^{\ell_1 \ell_2 \ell_3 \ell_4}_{T/E L}, \\
\alpha^{\ell_1 \ell_2 \ell_4 \ell_3}_{BL} &=& -(-1)^{L + \ell_{34}} ~\alpha^{\ell_1 \ell_2 \ell_3 \ell_4}_{B L},\\
\beta_{T/EL}^{\ell_1 \ell_2 \ell_4 \ell_3} &=& \gamma_{T/EL}^{\ell_1 \ell_2 \ell_3 \ell_4} = \beta_{T/EL}^{\ell_2 \ell_1 \ell_3 \ell_4} ,\\
\beta_{BL}^{\ell_1 \ell_2 \ell_4 \ell_3} &=& -\gamma_{BL}^{\ell_1 \ell_2 \ell_3 \ell_4} = - \beta_{BL}^{\ell_2 \ell_1 \ell_3 \ell_4}
\earr
where $\ell_{34} \equiv \ell_3 + \ell_4$. These properties also apply to the tilded counterparts.  

Using Eq.~\eqref{eq:alph_gen}, we have computed all necessary coefficients and listed them in Table \ref{Tab:coeffs} in Sec.~\ref{sec:trispec}.

\section{Explicit \texorpdfstring{$E$}{E}- and \texorpdfstring{$B$}{B}-mode power spectra calculations}\label{app:powerspec}

In this appendix we use the full formalism defined in Section~\ref{sec:trispec} to explicitly compute the several cross-power-spectra involving $E$- and $B$-modes. Because the perturbation to polarization is cubic in the primordial curvature perturbation, there are potential non-zero 2-point correlations with the standard unperturbed CMB anisotropy. The cross-correlation of $E$ and $\Theta$ with $B$ is zero due to parity considerations. However, albeit second-order in $f_{\rm pbh}$, the $B$ auto-correlation due to accreting PBHs is non-zero and has never been computed. We confirm the results in Section~\ref{subsec:CTE}, and additionally compute the $C^{(2)}_{BB\ell}$ power spectrum for the first time here.

Using Eqs.~\eqref{eq:X0_lm} and \eqref{eq:EB_transf_int}, we have
\begin{align}
    \langle E_{\ell m, \rm inh}^{(1)}& \Theta_{\ell' m'}^{*(0)} \rangle = \fbh \int\!\! D(k_1 k_2 k_3 k') ~   T^{(1)}_{E\ell m}(\bk_1, \bk_2, \bk_3) \nonumber\\
    &\quad\times T^{*(0)}_{\ell' m'}(\bk') \langle\zeta(\bk_1) \zeta(\bk_2) \zeta(\bk_3) \zeta^*(\bk') \rangle,\\
    \langle \Theta_{\ell m, \rm inh}^{(1)} &E_{\ell' m'}^{*(0)} \rangle = \fbh \int\!\! D(k_1 k_2 k_3 k') ~   T^{(1)}_{T\ell m}(\bk_1, \bk_2, \bk_3) \nonumber\\
    &\quad\times T^{*(0)}_{E\ell' m'}(\bk') \langle\zeta(\bk_1) \zeta(\bk_2) \zeta(\bk_3) \zeta^*(\bk') \rangle,\\
    \langle E_{\ell m, \rm inh}^{(1)}& E_{\ell' m'}^{*(0)} \rangle = \fbh \int\!\! D(k_1 k_2 k_3 k') ~   T^{(1)}_{E\ell m}(\bk_1, \bk_2, \bk_3) \nonumber\\
    &\quad\times T^{*(0)}_{E\ell' m'}(\bk') \langle\zeta(\bk_1) \zeta(\bk_2) \zeta(\bk_3) \zeta^*(\bk') \rangle,\\
    \langle B_{\ell m, \rm inh}^{(1)}& B_{\ell' m', \rm inh}^{*(1)} \rangle = \fbh^2 \int\!\! D(k_1 k_2 k_3 k_4 k_5 k_6) ~   \nonumber\\
    &\quad\langle\zeta(\bk_1) \zeta(\bk_2) \zeta(\bk_3)\zeta^*(\bk_4) \zeta^*(\bk_5) \zeta^*(\bk_6) \rangle \nonumber\\
    &\quad\quad\times T^{(1)}_{B\ell m}(\bk_1, \bk_2, \bk_3)  T^{*(1)}_{B\ell m}(\bk_4, \bk_5, \bk_6) .
    \end{align}
We solve these via Wick's theorem, and recalling that $T^{(1)}_{X\ell m}(\bk_1, - \bk_1, \bk_3) = 0$ where $X\in \{T,E,B\}$, and that $T^{(1)}_{X\ell m}$ is symmetric in its first two arguments.
\subsection{\texorpdfstring{$TE$}{TE} and \texorpdfstring{$EE$}{EE} power spectrum}
Focusing on the first equality, using Wick's theorem we have
\begin{align}\label{eq:E1T0}
   \langle E_{\ell m, \rm inh}^{(1)} \Theta_{\ell' m'}^{*(0)} \rangle= 2 \fbh  \int D(k k')   T^{(1)}_{E\ell m}(\bk', -\bk, \bk) \nonumber\\
\quad\times T^{*(0)}_{T\ell' m'}(\bk') P_{\zeta}(k) P_\zeta(k').
\end{align}
Taking a step back, if we write $T_{E\ell m}^{(1)}$ in terms of explicit helicity vectors, we have,
\begin{align}
T^{(1)}_{E\ell m}(\bk', -\bk, \bk) = -(\hk' \cdot \hk) \int_0^{\eta_0} \!\!\!d \eta ~g(\eta) \Delta_e(\eta, k') \Delta_e(\eta, k) \nonumber\\
 \times\frac{1}{2} \int d^2 \hn ~Y_{\ell m}^*(\hn)(\epsilon^i_{+}\epsilon^j_{+} +\epsilon^i_{-}\epsilon^j_{-})\widetilde{\mathcal{P}}^{(0)}_{ij}(\eta,\bk,\hn)  e^{-i \chi \hn \cdot \bk'},
\end{align}
When performing the ${\bm k}$ integral of Eq.~\eqref{eq:E1T0}, we see that we will have a resulting third rank tensor of the form,
\begin{align}
\mathcal{K}_{kij}(\eta, \hn)\equiv\int Dk~ \hk_k \widetilde{\mathcal{P}}^{(0)}_{ij}(\eta,\bk,\hn),
\end{align}
which can \textit{only} be constructed from combinations of $\hn$ vectors and Kronecker deltas due to integrating out the $\hk$ dependence. This implies $\epsilon^i_{\pm}\epsilon^j_{\pm}\mathcal{K}_{kij}=0$ because the helicity projections are orthogonal to each other and $\epsilon_\pm^i\hn_i=0$. Thus,
\begin{align}
    \langle E_{\ell m, \rm inh}^{(1)}\Theta_{\ell' m'}^{*(0)} \rangle=\langle E_{\ell m, \rm inh}^{(1)} E_{\ell' m'}^{*(0)} \rangle=0.
\end{align}
This confirms the general geometric considerations of Section~\ref{subsec:CTE}. For peace of mind, we also checked using the full formalism of the harmonic coefficients of the cubic transfer function, Eq.~\eqref{eq:T_multipoles}, and found the same results.

\subsection{\texorpdfstring{$B$}{B}-mode auto-power spectrum}
For the $B$-mode auto-correlation, considering the symmetries of $T^{(1)}_{\ell m}$ we will have terms
\begin{align}
     \langle B_{\ell m, \rm inh}^{(1)}&B_{\ell' m', \rm inh}^{*(1)} \rangle=\nonumber\\
    & \fbh^2  \int D(k k' k'') P_{\zeta}(k) P_\zeta(k')P_{\zeta}(k'')\nonumber\\    
    &\quad\times \left[4  T^{(1)}_{B\ell m}(\bk, \bk', -\bk) T^{*(1)}_{B\ell' m'}(\bk', -\bk'', \bk'')\nonumber\right.\\
    &\quad\quad\left.+2T^{(1)}_{B\ell m}(\bk, \bk', \bk'') T^{*(1)}_{B\ell' m'}(\bk, \bk', \bk'')\right.\nonumber\\
    &\quad\quad\left.+4T^{(1)}_{B\ell m}(\bk, \bk', \bk'') T^{*(1)}_{B\ell' m'}(\bk, \bk'', \bk')\right].
\end{align}
The first term is zero for the same reasons the $E$-mode cross-power spectra were zero above. We then have,
\begin{align}
&\langle B_{\ell m, \rm inh}^{(1)}B_{\ell' m', \rm inh}^{*(1)} \rangle=\nonumber\\
    &\quad  \fbh^2  \int D(k k'k'')(4 \pi)^3P_{\zeta}(k) P_\zeta(k')P_{\zeta}(k'') \!\!\!\!\!\!\sum_{\ell_1 \ell_2 \ell_3, m_1 m_2 m_3}\!\!\!\!\!\! \nonumber\\
    &\quad\quad\times\left[2T_{B\ell_1 \ell_2\ell_3; \ell}^{m_1 m_2 m_3; m}(k, k', k'')T_{B\ell_1 \ell_2\ell_3; \ell'}^{*m_1 m_2 m_3; m'}(k, k', k'')\right.\nonumber\\
    &\quad\quad\quad\!\!\left.+4T_{B\ell_1 \ell_2\ell_3; \ell}^{m_1 m_2 m_3; m}(k, k', k'')T_{B\ell_1 \ell_3\ell_2; \ell'}^{*m_1 m_3 m_2; m'}(k, k'', k')\right].
\end{align}
We exploit the fact that the Universe is statistically isotropic and compute, 
\begin{align}
    \langle B_{\ell m, \rm inh}^{(1)}B_{\ell' m', \rm inh}^{*(1)}\rangle=\frac{\delta_{\ell\ell'}\delta_{mm'}}{2\ell+1}\sum_{m''}\langle B_{\ell m'', \rm inh}^{(1)} B_{\ell m'', \rm inh}^{*(1)}\rangle.
\end{align}
We define 
\begin{align}
\langle B_{\ell m, \rm inh}^{(1)}B_{\ell' m', \rm inh}^{*(1)}\rangle=\delta_{\ell \ell'}\delta_{m m'}C^{(2)}_{BB\ell,\rm inh}.
\end{align}
Using similar methods from Appendix~E in Paper~II, when the dust settles we find,
\begin{align}\label{eq:inh_auto}
    C_{BB\ell,\rm inh}^{(2)}=f_{\rm pbh}^2\Bigr[\sum_{\ell_1\ell_2\ell_3}\left(2\mathfrak{B}_{\ell_1\ell_2,\ell_3;\ell}+4\mathfrak{C}_{\ell_1,\ell_2,\ell_3;\ell}\right)\Bigr],
\end{align}
where
\begin{align}
    &\mathfrak{B}_{\ell_1\ell_2,\ell_3;\ell}\equiv\frac{(4\pi)^3}{2\ell+1}\int d \eta\int d \eta' g(\eta)g(\eta')\nonumber\\
    &\times\left\{\mathcal{A}_{\ell_1}(\eta,\eta')\mathcal{A}_{\ell_2}(\eta,\eta') J_{\ell_3}(\eta,\eta')(\mathcal{Q}^2)_{\ell_1 \ell_2, \ell_3 \ell}\right.\nonumber\\
    &\quad+2\mathcal{K}_{\ell_1}(\eta,\eta')\mathcal{K}_{\ell_2}(\eta,\eta') J_{\ell_3}(\eta,\eta')(\mathcal{Q \widetilde{Q}})_{\ell_1 \ell_2, \ell_3 \ell}\nonumber\\
    &\quad\left.+\mathcal{B}_{\ell_1}(\eta,\eta')\mathcal{B}_{\ell_2}(\eta,\eta') J_{\ell_3}(\eta,\eta')(\mathcal{\widetilde{Q}}^2)_{\ell_1 \ell_2, \ell_3 \ell}\right\},
    \end{align}
    \begin{align}
    &\mathfrak{C}_{\ell_1,\ell_2,\ell_3;\ell}\equiv\frac{(4\pi)^3}{2\ell+1}\int d \eta \int d\eta' g(\eta)g(\eta')\nonumber\\
    &\times\left\{\mathcal{A}_{\ell_1}(\eta,\eta')\widetilde{\mathcal{A}}_{\ell_2} (\eta,\eta')\widetilde{\mathcal{A}}_{\ell_3}(\eta',\eta)(\mathcal{{Q} {Q}}^{\rm S})_{\ell_1, \ell_2 \ell_3, \ell}\right.\nonumber\\
    &\quad\!\!\!+2{\mathcal{K}}_{\ell_1}(\eta,\eta')\widetilde{\mathcal{A}}_{\ell_2}(\eta,\eta') \widetilde{\mathcal{B}}_{\ell_3}(\eta',\eta)(\mathcal{Q \widetilde{Q}}^{\rm S})_{\ell_1, \ell_2, \ell_3, \ell}\nonumber\\
    &\quad\!\!\!\left.+\mathcal{B}_{\ell_1}(\eta,\eta')\widetilde{\mathcal{B}}_{\ell_2}(\eta,\eta') \widetilde{\mathcal{B}}_{\ell_3}(\eta',\eta)(\mathcal{\widetilde{Q} \widetilde{Q}}^{\rm S})_{\ell_1, \ell_2 \ell_3, \ell} \right\}\!,
\end{align}
with
\begin{align*}
    \mathcal{A}_\ell(\eta,\eta')\equiv& \int\! Dk ~P_\zeta(k) \Delta_e(\eta, k) j_{\ell}'(k\chi)\Delta_e(\eta', k) j_{\ell}'(k\chi'),\nonumber\\
    \mathcal{B}_{\ell}(\eta,\eta')\equiv& f_\ell^2\!\! \!\int\!\! Dk~P_\zeta(k) \frac{j_{\ell}(\chi k)}{\chi k}\Delta_e (\eta, k)\frac{j_{\ell}(\chi' k)}{\chi' k}\Delta_e (\eta', k),\nonumber\\
    \mathcal{K}_{\ell}(\eta,\eta')\equiv& f_\ell\!\! \!\int\!\! Dk~ P_\zeta(k)j_{\ell}'(\chi k)\Delta_e (\eta, k)\frac{j_{\ell}(\chi' k)}{\chi' k}\Delta_e (\eta', k),\nonumber\\
    J_{\ell}(\eta,\eta')\equiv& \int Dk~ P_\zeta(k)\mathcal{J}_{E\ell}(\eta, k)\mathcal{J}_{E\ell}(\eta', k), \nonumber\\
    \widetilde{\mathcal{A}}_{\ell}(\eta,\eta')\equiv& \int\!\! Dk~ P_\zeta(k)j_{\ell}'(\chi k)\Delta_e (\eta, k)\mathcal{J}_{E\ell}(\eta', k),\nonumber\\
    \widetilde{\mathcal{B}}_{\ell}(\eta,\eta')\equiv& f_\ell \!\int\!\! Dk~P_\zeta(k) \frac{j_{\ell}(\chi k)}{\chi k}\Delta_e (\eta, k)\mathcal{J}_{E\ell}(\eta', k),\nonumber\\
    (\mathcal{Q}^2)_{\ell_1 \ell_2, \ell_3 \ell_4} \equiv& \sum_{m's} \left(Q_{B\ell_1 \ell_2, \ell_3 \ell_4}^{m_1 m_2, m_3 m_4}\right)^2,\\
    (\mathcal{Q \widetilde{Q}})_{\ell_1 \ell_2, \ell_3 \ell_4} \equiv& \sum_{m's} {Q^*}_{B\ell_1 \ell_2, \ell_3 \ell_4}^{m_1 m_2, m_3 m_4} \widetilde{Q}_{B\ell_1 \ell_2, \ell_3 \ell_4}^{m_1 m_2, m_3 m_4}, \\
 (\mathcal{\widetilde{Q}}^2)_{\ell_1 \ell_2, \ell_3 \ell_4} \equiv& \sum_{m's} \left( \widetilde{Q}_{B\ell_1 \ell_2, \ell_3 \ell_4}^{m_1 m_2, m_3 m_4}\right)^2, \\
 (\mathcal{{Q} {Q}}^{\rm S})_{\ell_1, \ell_2 \ell_3, \ell_4} \equiv& \sum_{m's} {Q^*}_{B\ell_1 \ell_2, \ell_3 \ell_4}^{m_1 m_2, m_3 m_4} {Q}_{B\ell_1 \ell_3, \ell_2 \ell_4}^{m_1 m_3, m_2 m_4},\\ 
 (\mathcal{{Q} \widetilde{Q}}^{\rm S})_{\ell_1, \ell_2, \ell_3, \ell_4} \equiv& \sum_{m's} {Q^*}_{B\ell_1 \ell_2, \ell_3 \ell_4}^{m_1 m_2, m_3 m_4} \widetilde{Q}_{B\ell_1 \ell_3, \ell_2 \ell_4}^{m_1 m_3, m_2 m_4},\\
 (\mathcal{\widetilde{Q} \widetilde{Q}}^{\rm S})_{\ell_1, \ell_2 \ell_3, \ell_4} \equiv& \sum_{m's} \widetilde{Q^*}_{B\ell_1 \ell_2, \ell_3 \ell_4}^{m_1 m_2, m_3 m_4} \widetilde{Q}_{B\ell_1 \ell_3, \ell_2 \ell_4}^{m_1 m_3, m_2 m_4}.
\end{align*}
 Using the terms discussed in Appendix~\ref{app:Q-sym} and defined in Table~\ref{Tab:coeffs}, and orthogonal properties of Wigner-$3j$ symbols, we can write the $Q$-symbols as (with 
 $g(L) \equiv 1/(2 L +1)$)
 \begin{align} \label{eq:Qstart}
(\mathcal{Q}^2)_{\ell_1 \ell_2 \ell_3 \ell_4} \equiv& \sum_{L}g(L) \left(\alpha^{\ell_1 \ell_2 \ell_3 \ell_4; B}_L\right)^2,\\
(\mathcal{Q \widetilde{Q}})_{\ell_1 \ell_2, \ell_3 \ell_4} \equiv& \sum_{L}g(L) \alpha^{*\ell_1 \ell_2 \ell_3 \ell_4; B}_L \widetilde{\alpha}^{\ell_1 \ell_2 \ell_3 \ell_4; B}_L, \\
 (\mathcal{\widetilde{Q}}^2)_{\ell_1 \ell_2, \ell_3 \ell_4} \equiv& \sum_{L}g(L) \left(\widetilde{\alpha}^{\ell_1 \ell_2 \ell_3 \ell_4; B}_L\right)^2, \\
 (\mathcal{{Q} {Q}}^{\rm S})_{\ell_1, \ell_2, \ell_3, \ell_4} \equiv& \sum_{L}g(L) {\alpha}^{*\ell_1 \ell_2 \ell_3 \ell_4; B}_L {\beta}^{\ell_1 \ell_3 \ell_2 \ell_4; B}_L,\\ 
 (\mathcal{{Q} \widetilde{Q}}^{\rm S})_{\ell_1, \ell_2, \ell_3, \ell_4} \equiv& \sum_{L}g(L) {\alpha}^{*\ell_1 \ell_2 \ell_3 \ell_4; B}_L \widetilde{\beta}^{\ell_1 \ell_3 \ell_2 \ell_4; B}_L,\\
 (\mathcal{\widetilde{Q} \widetilde{Q}}^{\rm S})_{\ell_1, \ell_2, \ell_3, \ell_4} \equiv& \sum_{L}g(L) \widetilde{\alpha}^{*\ell_1 \ell_2 \ell_3 \ell_4; B}_L \widetilde{\beta}^{\ell_1 \ell_3 \ell_2 \ell_4; B}_L.\label{eq:Qend}
\end{align}

\begin{figure}[htb]
\includegraphics[trim={0cm 1cm 0.5cm .5cm},width=.9\columnwidth]{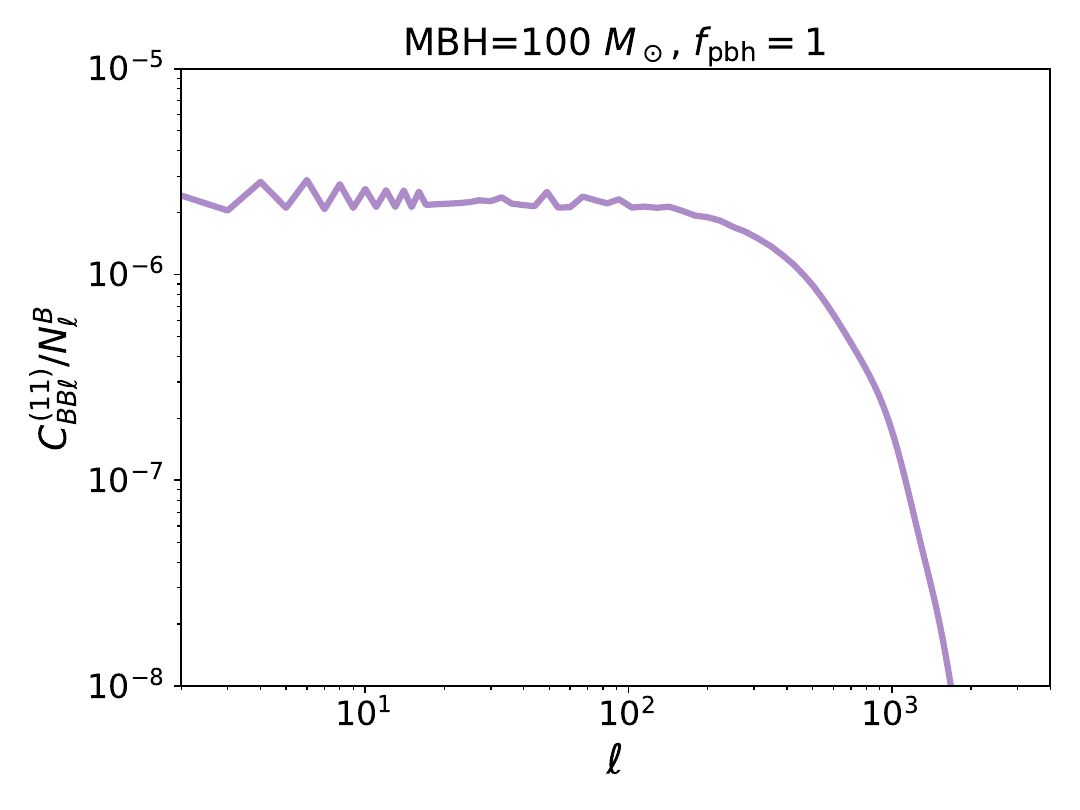}
\caption{\label{fig:auto} Auto-power-spectrum of the perturbed B-mode anisotropy due to accreting PBHs defined as $\langle B_{\ell m, \rm inh}^{(1)}B_{\ell' m', \rm inh}^{*(1)}\rangle=\delta_{\ell \ell'}\delta_{m m'}C^{(2)}_{BB\ell,\rm inh}$, normalized by the instrumental noise spectrum for polarization in a Planck-like experiment. We assume 100-$M_\odot$ PBHs comprising all the dark matter, but the qualitative trends are general for all PBH masses.}
\end{figure}

We plot the results, $C^{(11)}_{BB\ell,\rm inh}$, in Fig.~\ref{fig:auto}. We see this second-order effect (in $f_{\rm pbh}$) is heavily suppressed as compared to Planck's instrumental polarization noise spectrum (c.f. Sec.~\ref{subsec:noise}) by over five orders of magnitude. We also compare to the $BB$ power spectrum generated in \texttt{CLASS} from lensing, and find ours is suppressed by over three orders of magnitude for all $\ell$, indicating the effect from PBHs on $B$-mode 2-point statistics is negligible and not detectable. 

\section{Computational strategies}\label{app:comp}
In this appendix we describe the numerical methods and sampling used to compute the main result of this paper, the inverse variance of the estimator $\widehat{f}_{\rm pbh}$ from the temperature and polarization trispectra, Eq.~\eqref{eq:invvar}. For all relevant integrals involving $\eta, k$ and $\ell$, we use a similar sampling scheme as in Paper~II, which we summarize here.

Each conformal-time integral computed in this paper involves the visibility function $g(\eta)$, implying most of the support in the integrals comes from the recombination era. Starting sufficiently before recombination, $z_{\rm max} = 1400$, we sample $\eta$ with logarithmic step size $\Delta\ln \eta = 10^{-3}$ until $z_{\rm rec} = 900$, after which we increase the step size to $\Delta \ln \eta = 2\times 10^{-2}$ until $z_{\rm re} = 10$. From $z_{\rm re}$ until today, we sample linearly in $\eta$ with step size $\Delta\eta = 50$ Mpc.

Following in the same vein as Ref.~\cite{smith15a}, for $k$ integrals, we compute quantities on a grid using logarithmic spacing for low-$k$ and linear spacing at high-$k$. That is, we sample the wave number with step size $\Delta k={\rm min}(\epsilon k, \kappa_0)$ from $k_{\rm min}=10^{-5}$ Mpc$^{-1}$ to $k_{\rm max} = 5000 \eta_0^{-1} \approx 0.35$ Mpc$^{-1}$, where $\epsilon = 0.006$ and $\kappa_0 = 10^{-4}$ Mpc$^{-1}$.

Finally, we sample our $\ell$'s similar to the sampling in the standard output of \texttt{CLASS}. Our $\ell$ sampling consists of the floors of an array of real $\ell$ values spaced logarithmically in $2 \leq \ell < 400$ with $\Delta \ln \ell = 0.095$, and linearly in $400 \leq \ell < \ell_{\max} = 3000$ with $\Delta \ell = 39$. Note that in Paper~II, we doubled this resolution to check for convergence with regards to the temperature-only trispectrum. Because the polarization trispectra are more computationally intensive, we instead halved the resolution and found a fractional change of less than $10\%$.

With this sampling scheme, we compute and pretabulate $\mathcal{A}_{\ell_1\ell_2\ell_3}^{wxyz}$ and  $\widetilde{\mathcal{A}}_{\ell_1\ell_2\ell_3}^{wxyz}$ defined in Eq.~\eqref{eq:mathcalA} and Eq.~\eqref{eq:mathcalB} respectively. Because the transfer functions used to compute these quantities for either $E$ or $B$ are identical, we need only compute 16 three-dimensional $\ell$ tables for $\mathcal{A}^{wxyz}_{\ell_1\ell_2\ell_3}$ and $\widetilde{\mathcal{A}}^{wxyz}_{\ell_1\ell_2\ell_3}$ for every field configuration of $T,E,B$. We index the 16 tables by converting the four-digit number into binary where for $w,x,y,z \in \{T,E,B\}$, $T=0$ and $E,B=1$ (e.g. the field configuration $ETTB\rightarrow 1001$ is indexed as 9).

Inside the reduced trispectrum, ${T}^{w x y z}_{\ell_1 \ell_2 \ell_3 \ell_4}(L)$ defined in Eq.~\eqref{eq:red_tri}, exist five-dimensional quantities $\alpha$, $\beta$, and $\gamma$ defined in Appendix~\ref{app:Q-sym}. These quantities are unfeasible to store and we compute them in real-time in C code (Cython code to be precise). We do, however, precompute the 7 unique three-dimensional Wigner-3j symbols that comprise them.

With the $\mathcal{A}$, $\widetilde{\mathcal{A}}$ and Wigner-3j pretabulated, we compute $\sigma^2_{f_{\rm pbh}}$ via Eq.~\eqref{eq:invvar} by summing the products of reduced trispectra defined in Eq.~\eqref{eq:red_tri} over 7 for-loops. The first two loops are over every field configuration $a,b,c,d$ and $w,x,y,z$ for each of the two multiplied reduced trispectra; four of the loops iterate over $\ell_1\le \ell_2\le\ell_3\le\ell_4$; and finally the last loop over $L$. For the $\ell$'s we use the sampling discussed above. Because $\alpha$, $\beta$, and $\gamma$ are highly oscillatory, we sum over every single $0\le L\le 2\ell_{\rm max}$ (any $L$ sampled above $2\ell_{\rm max}$ results in a zero due to the triangle inequality property of Wigner-3j symbols) such that the resulting four-dimensional summation in $\ell$'s is sufficiently smooth to be approximated by an integral.

We modularize our code to deal with $\sum_{\ell_i}=\ell_1+\ell_2+\ell_3+\ell_4$ even or odd separately, and further more for $\ell_1+\ell_2+L$ even or odd separately. This is because the five dimensional quantities $\alpha$, $\beta$, and $\gamma$ are sparse depending on which fields we are considering. The sparsity is described by the following tree.\\

If $\sum_{\ell_i}$ is even:
\begin{itemize}
\item All $B$-related coefficients vanish. 
\item If $\ell_1 + \ell_2 + L$ is even: \vspace{-.15cm} 
    \begin{itemize}
        \item All $T$ and $E$ coefficients are nonzero.
    \end{itemize}
    \vspace{-.25cm}
\item If $\ell_1 + \ell_2 + L$ is odd:\vspace{-.15cm}
    \begin{itemize}
        \item  $\widetilde{\beta}$ and $\widetilde{\gamma}$ for both $E$ and $T$ are the only nonzero coefficients.
    \end{itemize}
\end{itemize}

If $\sum_{\ell_i}$ is odd:
\begin{itemize}
\item all $T$- and $E$-related coefficients vanish.
\item If $\ell_1 + \ell_2 + L$ is even: \vspace{-.15cm} 
    \begin{itemize}
        \item All $B$ coefficients are nonzero.
    \end{itemize}
    \vspace{-.25cm}
\item If $\ell_1 + \ell_2 + L$ is odd:\vspace{-.15cm}
    \begin{itemize}
        \item  $\beta$, $\widetilde{\beta}$ $\gamma$, and $\widetilde{\gamma}$ for $B$ are the only nonzero coefficients.
    \end{itemize}
\end{itemize}

To cross-check our results we also developed an independent code from the one above. Instead of placing the workload entirely on real-time CPU calculations, we instead load in the data ($\mathcal{A}$, $\widetilde{\mathcal{A}}$, and Wigner-3j symbols) in batches and utilize matrix multiplication on multiple GPUs. That is, this code uses vectorized operations similar to the \texttt{NumPy} Python package, but in the \texttt{PyTorch} framework with CUDA support. This enables us to write the code purely in Python with much better readability while still being feasibly fast. We found that both codes produced identical numerical results.

\end{appendix}

\newpage
\bibliography{main}

\end{document}